 \documentclass[12pt,journal,onecolumn]{IEEEtran}  

\usepackage{latexsym}
\usepackage{graphicx}
\usepackage{amsmath}
\usepackage{amssymb}
\usepackage{psfrag}
\usepackage{color}
\usepackage{cite}
\usepackage{times}
\usepackage{setspace}
\usepackage{multirow}
\usepackage{url}
\usepackage{flushend}
\usepackage[margin=1.1in]{geometry}
\usepackage{fancyhdr}

\newcommand{\stxt}[1]{\ensuremath{_{\text{#1}}}}
\renewcommand{\vec}[1]{{\mathbf{#1}}}

\newcommand{\chaina}{\mathcal{C}(3,6,L)}
\newcommand{\chainb}{\mathcal{C}(4,8,L)}

\begin{document}
\title{New Codes on Graphs Constructed by Connecting Spatially Coupled Chains}


\author{Dmitri Truhachev,~\IEEEmembership{Member,~IEEE,} David G. M. Mitchell,~\IEEEmembership{Member,~IEEE,} \\Michael~Lentmaier,~\IEEEmembership{Senior Member,~IEEE,}~and Daniel~J.~Costello,~Jr.,~\IEEEmembership{Life~Fellow,~IEEE}
\thanks{This work was supported in part by the National Science Foundation under Grant Number CCF-$1161754$ and in part by TELUS Corporation, Canada. The material in this paper was presented in part at the Information Theory and Applications Workshop, San Diego, CA, February 2012, in part at the IEEE International Conference on Communications, Ottawa, Canada, June 2012,  and in part at the IEEE International Symposium on Information Theory, Boston, MA, July 2012.}
\thanks{D.~Truhachev is with the Department of Electrical and Computer Engineering, Dalhousie University, Halifax, Canada (e-mail: dmitry@dal.ca)}
\thanks{D.~G.~M.~Mitchell and D.~J.~Costello,~Jr. are with the Department
of Electrical Engineering, University of Notre Dame, Notre Dame,
IN 46556, USA (e-mail: david.mitchell@nd.edu;~costello.2@nd.edu).}
\thanks{M. Lentmaier is with the Department of Electrical and Information Technology, Lund University, Lund, Sweden (e-mail: Michael.Lentmaier@eit.lth.se)}
}

\maketitle

\begin{abstract}
A novel code construction based on spatially coupled low-density parity-check (SC-LDPC) codes is presented. The proposed code ensembles are described by protographs, comprised of several protograph-based chains characterizing individual SC-LDPC codes. We demonstrate that code ensembles obtained by connecting appropriately chosen SC-LDPC code chains at specific points have improved iterative decoding thresholds compared to those of single SC-LDPC coupled chains. In addition, it is shown that the improved decoding properties of the connected ensembles result in reduced decoding complexity required to achieve a specific bit error probability. The constructed ensembles   are also asymptotically good, in the sense that the minimum distance grows linearly with the block length. Finally, we  show that the improved asymptotic properties  of the connected chain ensembles also translate into improved finite length performance. 
\end{abstract}

 
\section{Introduction}

Low-density parity-check  block codes (LDPC-BCs), invented by Gallager in the 1960's~\cite{Gal63} and later rediscovered in the 1990's~\cite{mac98,mac99}, have attracted a lot of attention recently in both the communications research community and for  telecommunication standards development due to their outstanding performance characteristics. However, the modern, low-complexity iterative decoding techniques generally employed for LDPC decoding are inferior to optimal maximum a posteriori probability (MAP) decoding, which is prohibitively complex for the operational lengths typical of LDPC codes. As a result, the limits of iterative belief propagation (BP) decoding (iterative decoding thresholds) of LDPC-BCs are below their MAP decoding thresholds~\cite{ru01b}. Achieving the MAP decoding threshold of LDPC codes is a desirable target, since it would bring communication rates even closer to the ultimate channel capacity limits.

To this end, it has been shown that the asymptotic iterative decoding performance of LDPC convolutional codes ($\mbox{LDPC-CC}$s), proposed in~\cite{fz99}, also called spatially coupled LDPC (SC-LDPC) codes, coincides with the optimal MAP decoding performance of underlying LDPC-BCs~\cite{lscz10,kru11,kru12,kymp12}. The explanation for this behavior is the phenomenon of spatial graph coupling that defines the structure of SC-LDPC codes. The principle of spatial graph coupling works in the following way. The Tanner graph of an initial small block code is replicated a number of times to produce a sequence (chain) of identical graphs. The neighboring copies of the initial graph are then connected by a set of edges. As a result, parity check nodes in the graph copies located at the boundaries of the chain are connected to a smaller number of variable nodes, and the groups of nodes at the ends of the chain form stronger sub-graphs with better protected variable nodes. As iterative message-passing decoding progresses, the nodes of the stronger sub-graphs at the chain boundaries generate more reliable information, which propagates through the chain from iteration to iteration. Remarkably, it has been proven that the iterative decoding thresholds of such SC-LDPC code ensembles coincide with the MAP decoding thresholds of the underlying LDPC-BC ensembles, which can be arbitrary close to the Shannon limit, for suitably large graph degree profiles. The principle of spatial graph coupling has attracted significant attention and has been successfully applied in many other areas of communications and signal processing~\cite{kp10,tru12comlet,tru13isit,ttk11,hkis11,sts11,ruas11,ypn11,au11}.

In this work, we demonstrate that graph coupling need not be limited to simply connecting component graphs into a single chain. Indeed, the coupling principle can be extended to more general structures. In particular, we propose novel protograph-based ensembles, in which  we construct new codes by connecting together several individual SC-LDPC code chains. As examples, we consider two ensembles obtained by connecting $(3,6)$-regular SC-LDPC code chains of various lengths and demonstrate that the chain connection results in improved iterative decoding thresholds on the binary erasure channel (BEC) and the additive white Gaussian noise (AWGN) channel compared to individual $(3,6)$-regular SC-LDPC chains of the same rate. In addition, we show that, like the component SC-LDPC chains (hereafter referred to as ``single chain ensembles''), the connected SC-LDPC code ensembles (hereafter simply ``connected chain ensembles'') are asymptotically good in the sense that the minimum distance grows linearly with the block length. We then consider several decoding schedules and show that the decoding complexity required to achieve specific decoding error probabilities in the near threshold region for the connected chain ensembles is also substantially reduced. Note that there are several degrees of freedom in the construction of connected chain ensembles: the types of codes to be connected, the lengths of the component chains, the connection point positions, and the structure of the connections all play important roles in determining  the decoding characteristics and resulting performance of the connected chain ensemble. We analyze the behavior of connected chain ensemble decoding and give insights into connected chain ensemble design using the above parameters. We also consider finite length codes taken from the constructed ensembles and show that their simulated error probability performance is superior to the codes obtained from single chain ensembles of the same rate and length. Finally, we consider  the connection of chains of different types and rates, including coupled irregular protograph-based  ARJA and AR4JA codes \cite{ddja09}, and show that similar improvements are also observed by connecting chains of these different types. 

The paper is organized as follows.  We start with an introduction to protograph-based  ensembles, describe regular single chain ensembles, and present two connected chain ensemble constructions, the ``square'' ensemble and the ``loop'' ensemble, in Section~\ref{sec:construction}. Section~\ref{sec:analysis} is dedicated to the analysis of the performance of connected chain ensembles. First, we consider transmission over the BEC, study the evolution of the decoding error probability at protograph nodes, and explain the improvements delivered by the connected chain ensembles. We then present results on iterative decoding thresholds for communication over the BEC and the AWGN channel in Section~\ref{sec:itthr}. Section~\ref{sec:complex} discusses  results on decoding complexity, and bit error rate (BER) simulation results for finite-length codes are given in Section~\ref{sec:sims}. Then results on minimum distance growth rates for the connected ensembles are described in Section~\ref{sec:distance}. Section~\ref{sec:dof} discusses several important features of the proposed constructions, such as choosing chain connection point positions and edge placement, mixing chains with different rates, thresholds, and graph densities, and connecting chains based on more general types of LDPC codes. Finally, some conclusions are  given in Section~\ref{sec:conc}.
\section{Code Construction}
\label{sec:construction}

We start by considering a single chain SC-LDPC code ensemble. Without loss of generality, we demonstrate our approach on an ensemble of coupled $(3,6)$-regular LDPC codes, constructed by means of protographs~\cite{tho03}. A protograph representing an LDPC code ensemble is a small bipartite graph connecting a set of variable nodes to a set of parity check nodes. Note that a protograph is different from the Tanner graph of a particular LDPC code, since every node of a protograph represents a set of $M$ nodes in the Tanner graph of a particular code and every edge represents a set of $M$ edges. The individual codes (members of the ensemble) are obtained via all possible permutations of these $M$ edges. As such, they are represented by the same protograph. Therefore, a protograph with a {\em lifting factor} of $M$ describes an ensemble of LDPC codes. It is an important feature of this construction that each lifted code inherits the degree distribution and graph neighbourhood structure of the protograph.

\begin{figure}[h]
\begin{center}
\setlength{\unitlength}{1mm}
   \begin{picture}(155,70)
  \put(0,0){\includegraphics{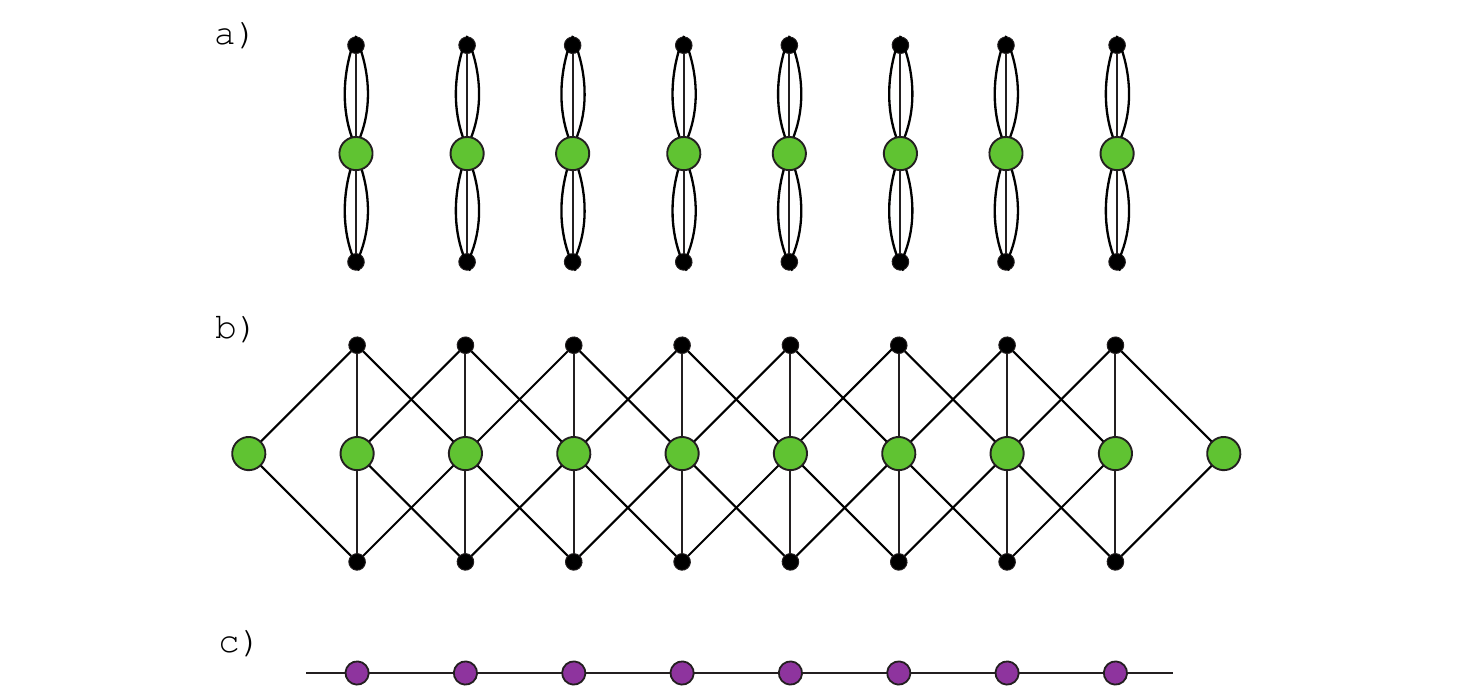}}
\end{picture}
\end{center}
\caption{Tanner graphs associated with (a) a chain of $L$ uncoupled $(3,6)$-regular LDPC-BC protographs for $L=8$ and (b) a single chain of $L$ spatially coupled $(3,6)$-regular LDPC-BC protographs for $L=8$ . The larger green circles in the figure correspond to parity-check nodes and the black circles correspond to variable nodes.  Also shown in (c) is a simplified illustration of the $(3,6)$-regular single chain protograph of length $L=8$. Each magenta node illustrates a segment consisting of one check node and two variable nodes.}
\label{Fig:parallel}
\end{figure}

A single chain SC-LDPC ensemble can be constructed by \emph{coupling} together several LDPC-BC ensembles into a chain. Fig.~\ref{Fig:parallel} shows representative Tanner graphs for (a) a group of uncoupled $(3,6)$-regular LDPC-BC ensemble protographs, (b) a single chain SC-LDPC ensemble protograph, and (c) a simplified illustration of the single chain protograph. As a result of the noninteracting structure, each component of the $(3,6)$-regular ensembles depicted in Fig.~\ref{Fig:parallel}(a) behaves independently, and we find the iterative BP decoding threshold for each protograph over the BEC is $\epsilon_{BP} = 0.4294$. By coupling together the $(3,6)$-regular LDPC-BC protographs, as demonstrated in Fig. \ref{Fig:parallel}(b) for $L=8$, we obtain the protograph of a single chain SC-LDPC ensemble. Note that, by coupling the block code protographs in this way, we introduce a ``structured irregularity'' into the coupled protograph. In this example all of the variable nodes still have $3$ edge connections: however, the check nodes at the start and the end of the chain are only connected to either $2$ or $4$ variable nodes. For this $(3,6)$-regular single chain SC-LDPC code ensemble, we find that the {\it threshold saturation} effect improves the BP threshold from the uncoupled BP threshold $\epsilon_{BP} = 0.4294$ to a value numerically indistinguishable from the (optimal) MAP threshold $\epsilon_{MAP} = 0.4881$ as the coupling length $L$ becomes sufficiently large~\cite{lscz10,kru11}.

The associated incidence matrix $\mathbf{B}$ of the protograph presented in Fig.~\ref{Fig:parallel}(b) is called the \emph{base} matrix and is given by

{\small
\[
\vec{B} = \left[
                  \begin{array}{cccccccccccccccc}
                    1\! & 1\! & 0\! & 0\! & 0\! & 0\! & 0\! & 0\! & 0\! & 0\! & 0\! & 0\! & 0\! & 0\! & 0\! & 0  \\
                    1\! & 1\! & 1\! & 1\! & 0\! & 0\! & 0\! & 0\! & 0\! & 0\! & 0\! & 0\! & 0\! & 0\! & 0\! & 0  \\
                    1\! & 1\! & 1\! & 1\! & 1\! & 1\! & 0\! & 0\! & 0\! & 0\! & 0\! & 0\! & 0\! & 0\! & 0\! & 0  \\
                    0\! & 0\! & 1\! & 1\! & 1\! & 1\! & 1\! & 1\! & 0\! & 0\! & 0\! & 0\! & 0\! & 0\! & 0\! & 0  \\
                    0\! & 0\! & 0\! & 0\! & 1\! & 1\! & 1\! & 1\! & 1\! & 1\! & 0\! & 0\! & 0\! & 0\! & 0\! & 0  \\
                    0\! & 0\! & 0\! & 0\! & 0\! & 0\! & 1\! & 1\! & 1\! & 1\! & 1\! & 1\! & 0\! & 0\! & 0\! & 0  \\
                    0\! & 0\! & 0\! & 0\! & 0\! & 0\! & 0\! & 0\! & 1\! & 1\! & 1\! & 1\! & 1\! & 1\! & 0\! & 0  \\
                    0\! & 0\! & 0\! & 0\! & 0\! & 0\! & 0\! & 0\! & 0\! & 0\! & 1\! & 1\! & 1\! & 1\! & 1\! & 1  \\
                    0\! & 0\! & 0\! & 0\! & 0\! & 0\! & 0\! & 0\! & 0\! & 0\! & 0\! & 0\! & 1\! & 1\! & 1\! & 1  \\
                    0\! & 0\! & 0\! & 0\! & 0\! & 0\! & 0\! & 0\! & 0\! & 0\! & 0\! & 0\! & 0\! & 0\! & 1\! & 1  \\
                  \end{array}
\right].
\]}
\normalsize

\noindent The parity-check matrix $\mathbf{H}$ of a protograph-based LDPC-BC can be created by replacing each non-zero entry $B_{i,j}$ in $\mathbf{B}$  by a sum of $ B_{i,j}$  permutation matrices of size $M$ and each zero entry by the $M\times M$ all-zero matrix. In graphical terms, this can be viewed as taking an $M$-fold graph cover or ``lifting'' of the protograph.  We denote the $(3,6)$-regular single chain ensemble protograph of length $L$ by $\mathcal{C}(3,6,L)$. The design rate of the ensemble $\mathcal{C}(3,6,L)$ is given by\footnote{Here $R$ denotes the design rate of the ensembles. The actual rate of a particular member of the ensemble may be slightly higher due to possible linear dependencies between the rows in its parity-check matrix.}
\begin{equation}
R(\mathcal{C}(3,6,L)) = \frac{L-2}{2L},
\end{equation}
which increases monotonically with $L$ and approaches $1/2$ as $L\to\infty$.

\subsection{Two Chains Connected by Bridges (The Square)}

\begin{figure}[h]
\setlength{\unitlength}{1mm}
   \begin{picture}(155,85)
  \put(0,0){\includegraphics{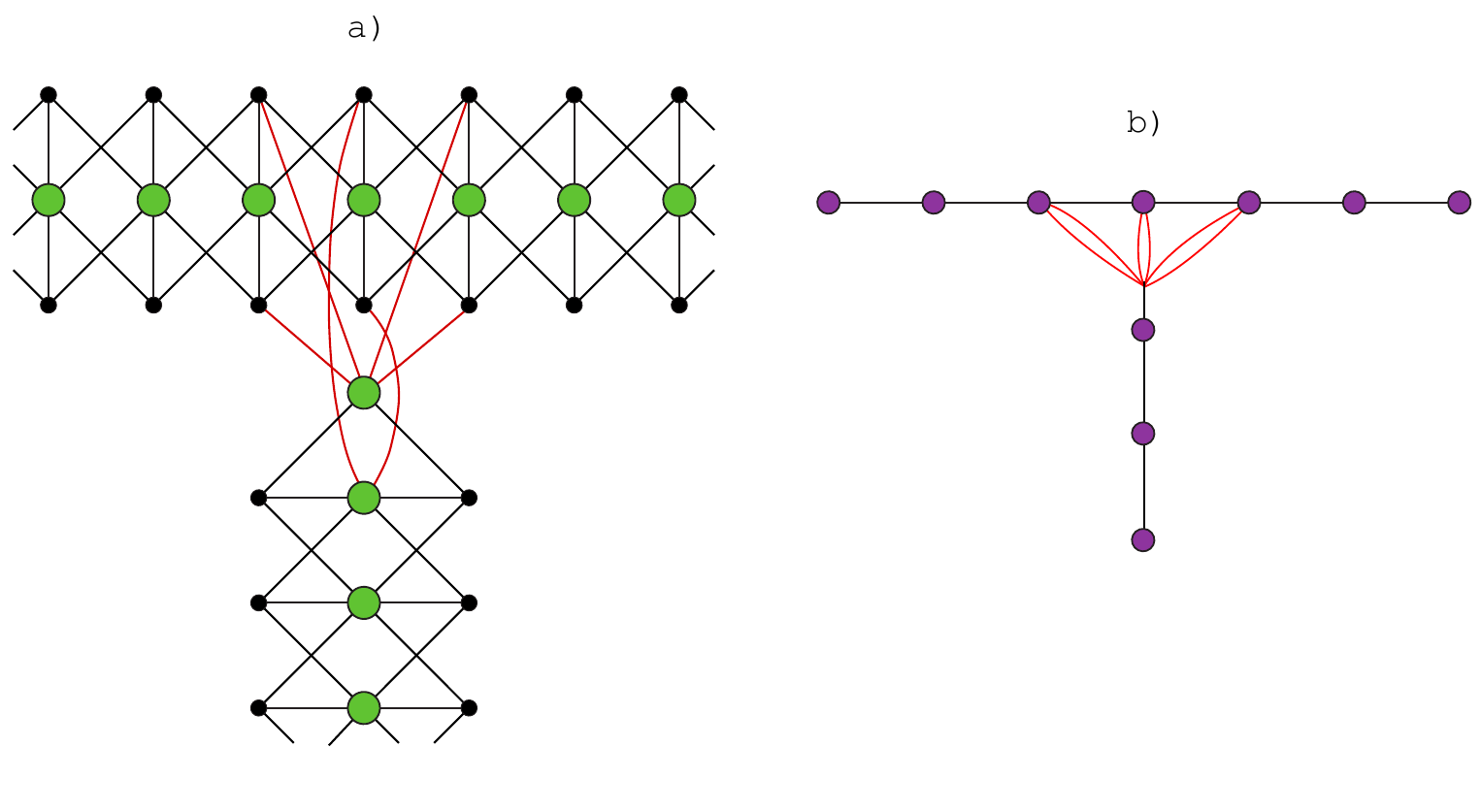}
   }
\end{picture}
\caption{Two connected $(3,6)$-regular single chain protographs. The connecting edges are shown in red. 
}
\label{Fig:connector}
\end{figure}

Two single chain protographs may be connected by adding edges between the check and variable nodes of the two protographs. Fig.~\ref{Fig:connector} shows two connected $(3,6)$-regular single chain protographs. 
Recall that, due to the boundary effects of coupling, the first check node at the top of the vertical chain has degree two and the second check node has degree four. In this example, the connections between the chains are chosen so that the check nodes at the end of the vertical chain have degree six. Using four additional edges, the first check node in the vertical chain is connected to variable nodes in the horizontal chain, and two additional edges are used to connect the second check node in the vertical chain to variable nodes in the horizontal chain. As a result, the degrees of six variable nodes in the horizontal chain are increased by one. The additional edges connecting the chains are shown in red in Fig.~\ref{Fig:connector}(a), and the simplified representation of this connection is shown in Fig.~\ref{Fig:connector}(b).

\begin{figure}[h]
\setlength{\unitlength}{1mm}
   \begin{picture}(155,65)
   \put(0,0){\includegraphics{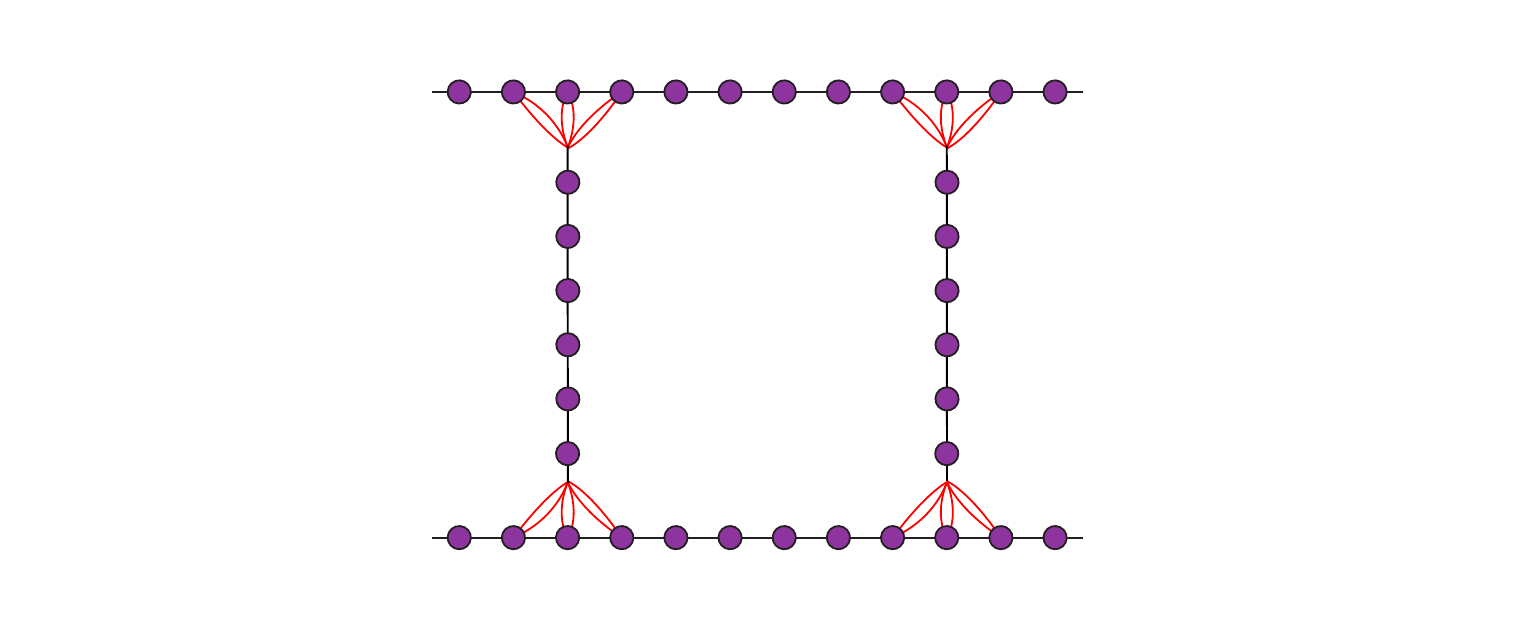}
   }
\end{picture}
\caption{Two single chain protographs of length $L=12$ connected by two bridges of length $L\stxt{b}=6$.}
\label{Fig:2c2b}
\end{figure}

We now consider two horizontal chains connected by two vertical chains that we call ``bridges''. An example of such a construction is given in Fig.~\ref{Fig:2c2b}. The length of both horizontal chains is $12$, while each bridge is of length $6$. Each of the connections between bridges and horizontal chains is made as shown in Fig.~\ref{Fig:connector}.

Generalizing this example, we consider a connected chain ensemble of protographs, denoted by $\mathcal{S}(3,6,L)$, that consists of two $(3,6)$-regular chains of length $L$ that are connected by two bridges of length $L/2$. The connecting points are located at a distance of $\lfloor L/4 \rfloor$ from the ends of the chains. The rate of the ensemble $\mathcal{S}(3,6,L)$ is given by
\begin{equation}
R(\mathcal{S}(3,6,L)) = \frac{3L-8}{6L} = \frac{1}{2}-\frac{4}{3L}\ .
\end{equation}


\subsection{Two Connected Chains (The Loop)}

\begin{figure}[h]
\setlength{\unitlength}{1mm}
   \begin{picture}(155,35)
   \put(0,0){\includegraphics{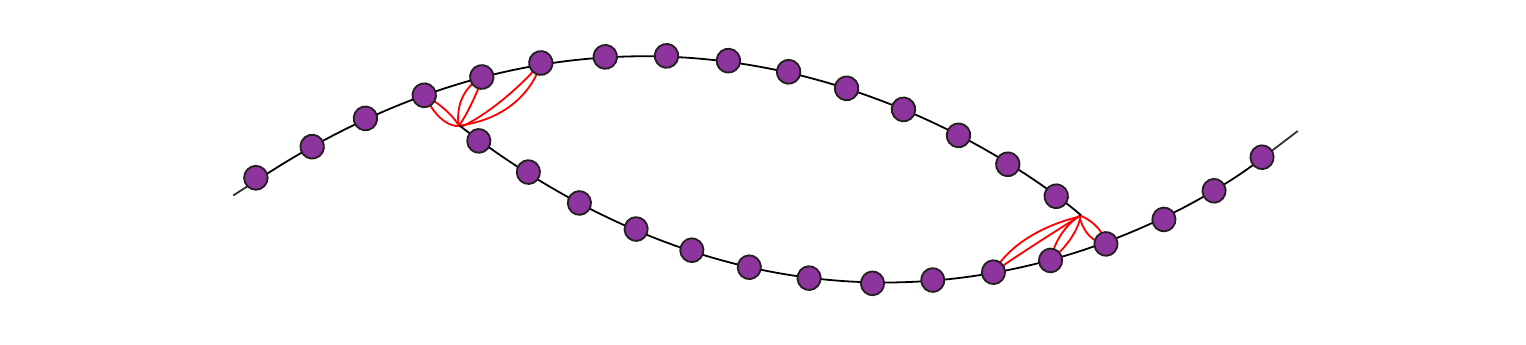}
   }
\end{picture}
\caption{Two single chain protographs of length $L=15$ connected as a loop.}
\label{Fig:loop}
\end{figure}

By directly connecting two single chain protographs (without bridges), we obtain the so-called ``loop'' ensemble. Fig.~\ref{Fig:loop} depicts the protograph consisting of two single chain protographs of length $L$ connected as a loop. Here, the last segment of the first chain is connected to an inner segment of the second chain, while the first segment of the second chain is connected to an inner segment of the first chain. The connections between the end of one chain and the inner part of the other are made as depicted in Fig.~\ref{Fig:connector}. The connecting points are located at a distance of $\lfloor L/3 \rfloor$ from the chain boundaries.  We denote the loop ensemble consisting of two $(3,6)$-regular single chains of length $L$ by $\mathcal{L}(3,6,L)$. Since the loop is constructed from two equal length chains, the rate of this ensemble is equal to the rate of a single chain
\begin{equation}
R(\mathcal{L}(3,6,L)) = R(\mathcal{C}(3,6,L)) = \frac{L-2}{2L}\ .
\end{equation}

\section{Analysis and Results}
\label{sec:analysis}

We start by studying the iterative decoding process for transmission over the BEC and compare the bit erasure probability evolution for nodes in both a loop and a single chain ensemble of the same rate. After that we present the iterative decoding threshold results for the connected chain ensembles and demonstrate that connected chain ensembles have a smaller decoding complexity in the near-threshold region. We then present some numerical finite length performance results and conclude this section with an analysis of the asymptotic minimum distance growth rates for the connected chain ensembles.  

\subsection{Iterative Decoding Convergence for Single and Connected Chain Ensembles}

Consider communication over a BEC with erasure probability $\epsilon$ using the  $\mathcal{L}(3,6,15)$ loop ensemble and $\mathcal{C}(3,6,15)$ single chain ensemble. We utilize density evolution to compute bit erasure probabilities at each node of the protograph for every decoding iteration. Using this tool we relate the evolution of the erasure probability to the node position in the protograph and compare the erasure probability behavior of the loop and chain ensembles. We explain the  relation of the bit erasure probability evolution to the ensemble structure and obtain insight into the performance improvement offered by connected chain ensembles. 

We denote the set of variable protograph nodes connected to check node $k$ in the protograph by $\mathbb{V}(k)$ and the set of check nodes connected to variable node $j$ by $\mathbb{C}(j)$. The probability that the message passed from check node $k$ to variable node $j$ at iteration $i$ is an erasure is denoted by $q_{kj}^{(i)}$. The probability of an erasure message from variable node $j$ to check node $k$ is similarly denoted by $p_{jk}^{(i)}$. The following equations relate the erasure probabilities of the messages at different iterations:
\begin{align}
q_{kj}^{(i)} &= 1 - \prod_{j' \in \mathbb{V}(k) \smallsetminus j} (1 - p_{j'k}^{(i-1)})\ , \\
p_{jk}^{(i)} &= \epsilon \prod_{k' \in \mathbb{C}(j) \smallsetminus k} q_{k'j}^{(i)}\ .
\end{align}
The variable node messages are initialized as $p_{jk}^{(0)}=\epsilon$ at iteration $0$. The bit erasure probability of the variable nodes at iteration $i$ can be calculated as
\begin{equation}
P\stxt{b}(j) = \epsilon \prod_{k \in \mathbb{C}(j) } q_{kj}^{(i)}\ .
\end{equation}

\begin{figure}[h]
\setlength{\unitlength}{1mm}
   \begin{picture}(155,90)
   \put(0,0){\includegraphics{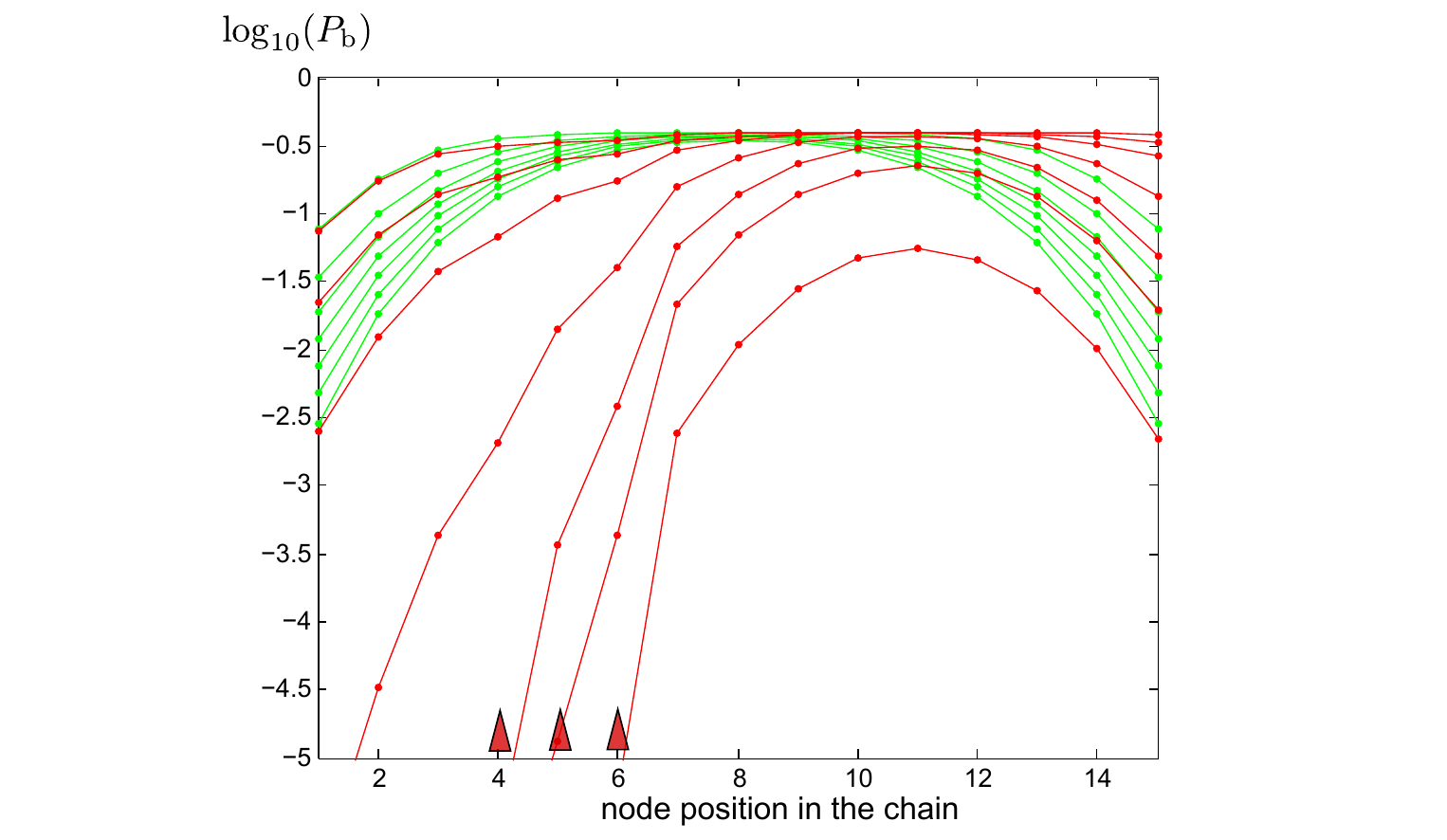}
   }
\end{picture}
\caption{Logarithm of the average bit erasure probability for the variable nodes of the first chain for the ensembles $\mathcal{L}(3,6,15)$ (red curves) and  $\mathcal{C}(3,6,15)$ (green curves), as a function of the position of the node in the chain. The curves (either red or green) are computed for decoding iterations $1,6,11,\ldots,36$ (from top to bottom). The three positions where the $1$st chain is connected to the end of the $2$nd chain in the loop are shown by the red triangles.}
\label{Fig:Pb_chain}
\end{figure}

The evolution of the bit erasure probability for the variable nodes of the $\mathcal{L}(3,6,15)$ ensemble is illustrated in Fig.~\ref{Fig:Pb_chain}. The red curves correspond to the erasure probability at each node position in chain one of the loop at iterations $1,6,11,\ldots,36$ (from top to bottom).\footnote{Due to the symmetric nature of the loop construction, it is sufficient to consider the evolution of the bit erasure probability for only one chain.} The green curves correspond to the erasure probability as a function of the node position for the single chain ensemble $\mathcal{C}(3,6,15)$ and iteration numbers $1,6,11,\ldots,36$. The BEC erasure probability is fixed to be $\epsilon = 0.488$. We notice that the red curves display lower error probability values with fewer iterations than the green curves, and hence it takes fewer decoding iterations for the ensemble $\mathcal{L}(3,6,15)$ to converge to a given bit erasure probability value.

Note that each green curve displays a perfect bell shaped figure and is concave. On the contrary, the shape of the red curves is not symmetric. This is due to the fact that the loop protograph is comprised of two connected chains. The inner part of the $1$st chain is connected to the $2$nd chain by edges connecting to nodes at positions $4$, $5$, and $6$, shown by red triangles on the figure. Note that the upper red curves dip down at these positions since the $2$nd chain provides convergence improvement via the connection. The  connection point creates a stronger sub-graph in which variable nodes are connected to four check nodes instead of three. This sub-graph distributes more reliable information to its neighborhood (similar to the open end (positions $1$ and $2$) of the first chain) throughout the decoding precess. 

On the other hand, the end of the first chain (positions $14$ and $15$) converges to low bit erasure probability values more slowly than the single chain. This can be observed by comparing to the corresponding green curves. The reason for this behavior is the additional connecting edges that are now present at the end of the $1$st chain and which connect it to the $2$nd. These edges are absent in the single chain case. However, convergence at the end of the 1st chain improves with subsequent iterations as the $2$nd chain starts to converge. Eventually, the lower red curve displays a perfect bell shape as a result of more reliable information coming from the second chain. 

The two connected chains create a balanced system in which one helps the other to converge and vice versa. The distances between the connection points as well as the positions of the edges connecting the two chains are important parameters. Their role will be discussed in Section~\ref{sec:dof}.


\subsection{Iterative Decoding Thresholds}
\label{sec:itthr}

The improved convergence behavior resulting from the balanced exchange of reliable information between the two chains implies improved robustness to channel noise. One of the consequences is improved iterative decoding thresholds of the connected chain ensembles. 

The BEC thresholds $\epsilon^*$ for the $\mathcal{L}(3,6,L)$ ensembles, where $L=12,15$, and $18$, are shown in Table~\ref{tab:thres}. The thresholds of
the single chain ensembles of the same rates are presented for comparison. It can be observed that the thresholds of the
connected ensembles are always better than the thresholds of the corresponding (equal rate) single chain ensembles. 

\begin{table}[h]
\begin{center}
\scalebox{1}{%
\begin{tabular}{|c|c|c|c|c|c|}\hline
Rate & Ensemble & $\epsilon^*$ & Ensemble & $\epsilon^*$ \\\hline
$0.4167$ & $\mathcal{L}(3,6,12)$ & $0.5237$ & $\mathcal{C}(3,6,12)$ & $0.495$\\
$0.4333$ & $\mathcal{L}(3,6,15)$ &  $0.5105$ & $\mathcal{C}(3,6,15)$ & $0.489$\\
$0.4444$ & $\mathcal{L}(3,6,18)$ &  $0.4989$ & $\mathcal{C}(3,6,18)$ &  $0.488$ \\
\hline
\end{tabular}}
\end{center}
\caption{BEC thresholds $\epsilon^*$ for several loop ensembles $\mathcal{L}(3,6,L)$ and single chain ensembles $\mathcal{C}(3,6,L)$.}\label{tab:thres}
\end{table}

AWGN channel thresholds for the loop ensembles $\mathcal{L}(3,6,L)$ are given in Table~\ref{tab:gauss} for $L=12,15,$ and $18$, and the results for the single chain ensembles $\mathcal{C}(3,6,L)$ are shown for comparison. Again, we notice that
the thresholds of the loop ensembles are significantly better than for the corresponding single chains.\vspace{-1mm}

\begin{table}[h]
\begin{center}
\scalebox{1}{%
\begin{tabular}{|c|c|c|c|c|c|c|}\hline
Rate & Ensemble & $(E\stxt{b}/N_0)^*$ & Ensemble & $(E\stxt{b}/N_0)^*$ \\\hline
$0.4167$ & $\mathcal{L}(3,6,12)$ & $0.6520$dB & $\mathcal{C}(3,6,12)$ & $1.1167$dB\\
$0.4333$ & $\mathcal{L}(3,6,15)$ &  $0.7281$dB & $\mathcal{C}(3,6,15)$ & $1.0431$dB\\
$0.4444$ & $\mathcal{L}(3,6,18)$ &  $0.7850$dB & $\mathcal{C}(3,6,18)$ & $0.9659$dB\\
\hline
\end{tabular}}
\end{center}
\caption{\label{tab:gauss}AWGN channel thresholds $(E\stxt{b}/N_0)^*$ calculated for the $\mathcal{L}(3,6,L)$ loop ensembles and the $\mathcal{C}(3,6,L)$ ensembles for $L=12$, $15$, and $18$.}\label{tab:thresAWGN}
\end{table}

\begin{figure}[h]
\setlength{\unitlength}{1mm}
   \begin{picture}(155,100)
   \put(0,0){\includegraphics{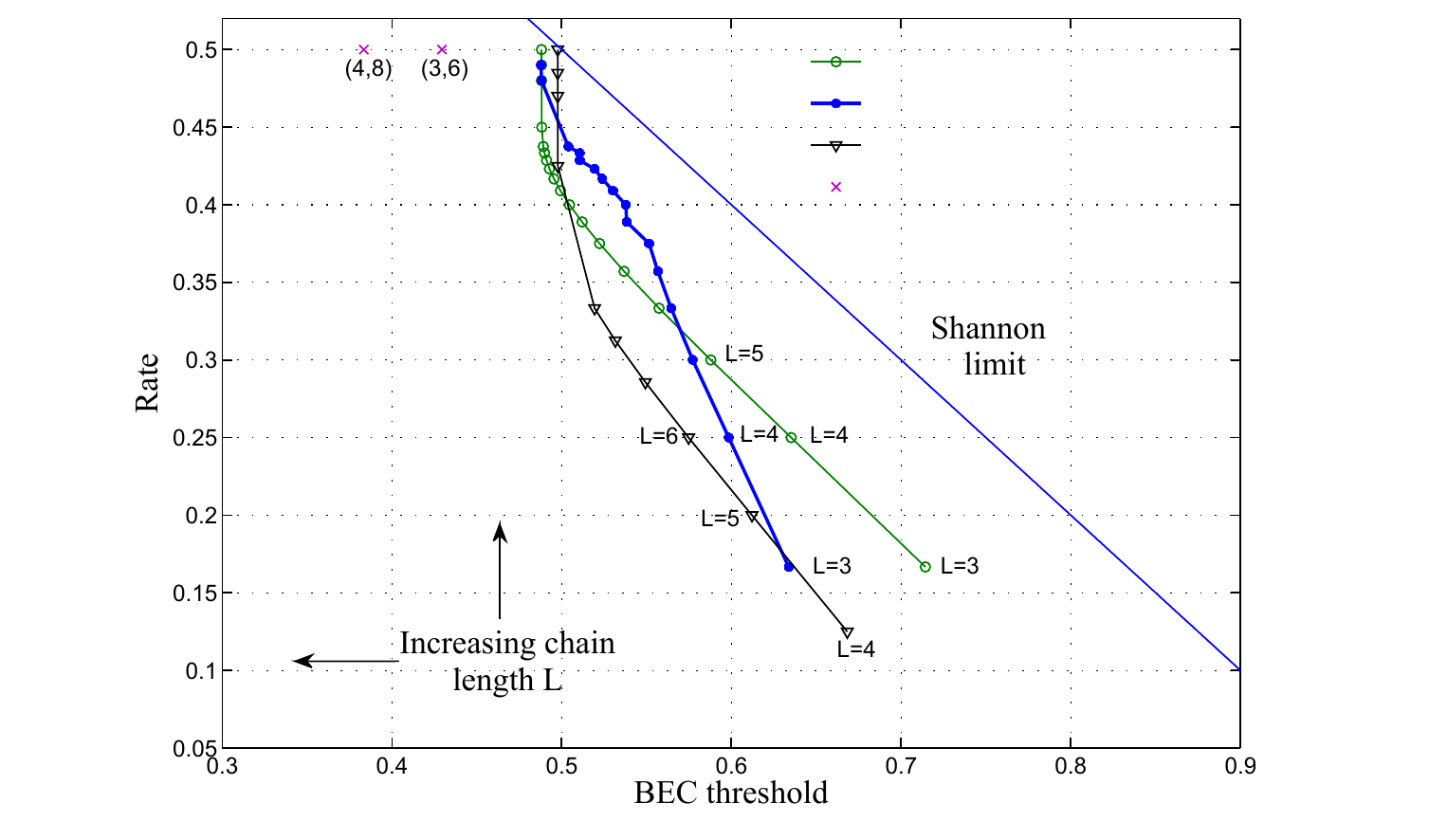}}
   \put(95,69){\small $(J,K)$-regular LDPC}
   \put(95,82){\small $\mathcal{C}(3,6,L)$}
   \put(95,77.5){\small $\mathcal{L}(3,6,L)$}
   \put(95,73){\small $\mathcal{C}(4,8,L)$}
\end{picture}
\caption{BEC thresholds for the $\mathcal{L}(3,6,L)$ loop ensembles as well as some $\mathcal{C}(J,K,L)$ ensembles and $(J,K)$-regular LDPC-BC ensembles.}
\label{fig:36thresh}
\end{figure}
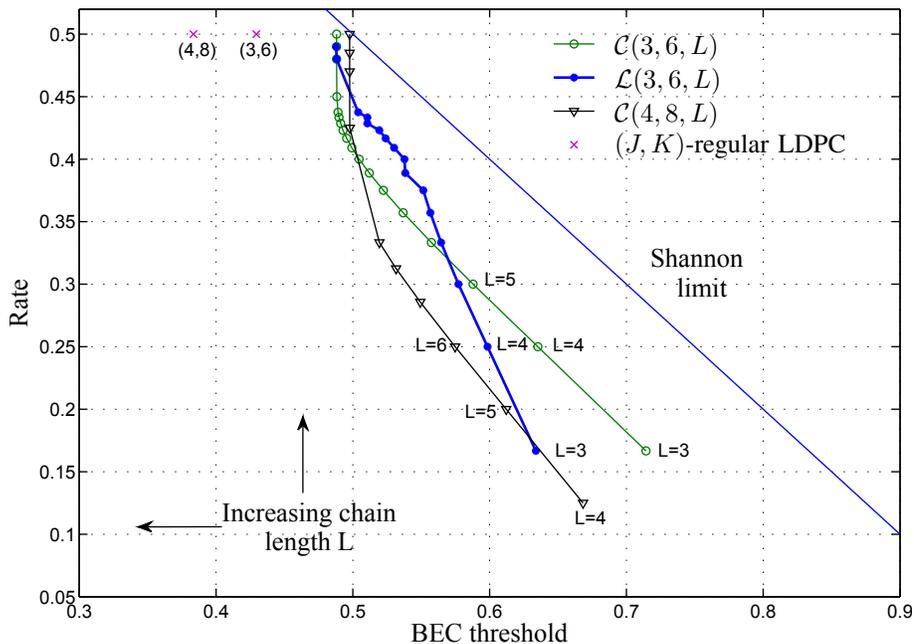

Fig.~\ref{fig:36thresh} shows the BEC thresholds for the $\mathcal{L}(3,6,L)$ ensembles in comparison to the $\mathcal{C}(3,6,L)$ and $\mathcal{C}(4,8,L)$ single chain  ensembles for a variety of chain lengths $L$. Comparing the $\mathcal{L}(3,6,L)$ ensembles to the  $\mathcal{C}(3,6,L)$ ensembles, we observe that, for $L>5$, the thresholds  of the loop ensembles are generally superior, with the exception of large $L$. In particular, we observe a dramatic threshold improvement for ensembles with rates in the region $0.35 \leq R_L \leq 0.45$. For larger values of $L$, the improvement diminishes. The thresholds of the single chain ensembles $\chaina$ and $\chainb$ are observed to converge to values close to the MAP threshold of the underlying $(3,6)$- and $(4,8)$-regular LDPC-BC ensembles as $L$ becomes sufficiently large. 
As a result, for large $L$, we observe the surprising behavior that the iterative decoding thresholds of the $\chainb$ ensembles are larger than the $\chaina$ thresholds (unlike the corresponding LDPC-BC ensembles). However, even in this region, we observe that the thresholds of the $\mathcal{L}(3,6,L)$ ensemble remain above the $\chainb$ thresholds for rates between $0.4$ and $0.45$. In the next section, we will see that a loop constructed using $(4,8)$-regular single chain protographs achieves further performance improvement.



\normalsize

The thresholds of the square ensembles also improve compared to the single chain ensembles. Iterative decoding thresholds for a number of $\mathcal{S}(3,6,L)$ ensembles on the BEC are given in Table~\ref{tab:thres}. The thresholds are compared with the thresholds of the regular coupled chains of (approximately) the same rate. The square ensemble thresholds for the AWGN channel, computed using a discretized density evolution method, are given in Table~\ref{tab:thresAWGN}. 

\begin{table}[h]
\begin{center}
\scalebox{1}{%
\begin{tabular}{|c|c|c|c|c|c|}\hline
Rate & Ensemble & $\epsilon^*$ & Ensemble & $\epsilon^*$ \\\hline
$0.3333$ & $\mathcal{S}(3,6,8)$ &  $0.563$ & $\mathcal{C}(3,6,6)$ & $0.557$\\
$0.3889$ & $\mathcal{S}(3,6,12)$ & $0.538$  & $\mathcal{C}(3,6,9)$ & $0.512$\\
$0.4167$ & $\mathcal{S}(3,6,16)$ & $0.522$ & $\mathcal{C}(3,6,12)$ & $0.495$\\
$0.4333$ & $\mathcal{S}(3,6,20)$ &  $0.504$ & $\mathcal{C}(3,6,15)$ & $0.489$\\
$0.4444$ & $\mathcal{S}(3,6,24)$ &  $0.495$ & $\mathcal{C}(3,6,18)$ &  $0.488$ \\
\hline
\end{tabular}}
\end{center}
\caption{BEC thresholds $\epsilon^*$ for several square ensembles $\mathcal{S}(3,6,L)$ and single chain ensembles.}\label{tab:thres}\vspace{-4mm}
\end{table}

\begin{table}[h]
\begin{center}
\scalebox{1}{%
\begin{tabular}{|c|c|c|c|c|c|}\hline
Rate & Ensemble & $(E\stxt{b}/N_0)^*$ & Ensemble & $(E\stxt{b}/N_0)^*$ \\\hline
$0.3333$ & $\mathcal{S}(3,6,8)$ &  $1.0167$dB & $\mathcal{C}(3,6,6)$ & $1.1894$dB\\
$0.3889$ & $\mathcal{S}(3,6,12)$ & $0.7512$dB  & $\mathcal{C}(3,6,9)$ & $1.1701$dB\\
$0.4167$ & $\mathcal{S}(3,6,16)$ & $0.7231$dB & $\mathcal{C}(3,6,12)$ & $1.1167$dB\\
$0.4333$ & $\mathcal{S}(3,6,20)$ &  $0.8079$dB & $\mathcal{C}(3,6,15)$ & $1.0431$dB\\
$0.4444$ & $\mathcal{S}(3,6,24)$ &  $0.8367$dB & $\mathcal{C}(3,6,18)$ & $0.9659$dB\\
\hline
\end{tabular}}
\end{center}
\caption{AWGN channel thresholds $(E\stxt{b}/N_0)^*$ for several square ensembles $\mathcal{S}(3,6,L)$ and single chain ensembles.}\label{tab:thresAWGN}\vspace{-4mm}
\end{table}

Finally, we consider higher rate ensembles constructed by connecting $(3,9)$-regular protograph chains. The BEC thresholds calculated for the ensembles $\mathcal{L}(3,9,L)$ and $\mathcal{C}(3,9,L)$ are given in Table~\ref{tab:39thresh}. We see that the thresholds of the $\mathcal{L}(3,9,L)$ ensemble are again improved compared to the thresholds of the $\mathcal{C}(3,9,L)$ ensemble for all  moderate values of $L$, but that the advantage disappears for large $L$.
\begin{table}[h]
\begin{center}
\begin{tabular}{|c|c|c|c|c|c|}\hline
Rate & Ensemble & $\epsilon^*$ & Ensemble & $\epsilon^*$ \\\hline
$0.5556$ & $\mathcal{L}(3,9,6)$ &  $0.3746$ & $\mathcal{C}(3,9,6)$ & $0.3605$\\
$0.5883$ & $\mathcal{L}(3,9,8)$ & $0.3604$  & $\mathcal{C}(3,9,8)$ & $0.3392$\\
$0.6111$ & $\mathcal{L}(3,9,12)$ & $0.3437$ & $\mathcal{C}(3,9,12)$ & $0.3235$\\
$0.6600$ & $\mathcal{L}(3,9,100)$ &  $0.3191$ & $\mathcal{C}(3,9,100)$ & $0.3196$\\
\hline
\end{tabular}
\end{center}
\caption{BEC thresholds for the loop ensemble $\mathcal{L}(3,9,L)$ and the single chain ensemble $\mathcal{C}(3,9,L)$.}\label{tab:39thresh}
\end{table}

\subsection{Decoding Complexity}
\label{sec:complex}


To focus on a reduction in decoding complexity, we consider simultaneous decoding of the entire code graph, where we employ the updating schedule proposed in~\cite{lpf11}. The algorithm designates a target bit erasure probability $P\stxt{b,max}$ as well as an update improvement parameter $\theta$. Regular message passing updates are performed for each variable or check node with the following exceptions:
\begin{itemize}
\item no update for variable node $j$ is performed if the bit erasure probability $P\stxt{b}(j) < P\stxt{b,max}$;
\item no update for any variable node $j$ or any check node $k$ is performed if all the nodes in $\mathbb{C}(j)$ or $\mathbb{V}(k)$, respectively, were not updated in the previous iteration;
\item no update for variable node $j$ is performed if the potential improvement of the bit erasure probability is less than $\theta$, i.e.,
\begin{equation}
\frac{P\stxt{b,old}(j) - P\stxt{b,new}(j)}{P\stxt{b,old}(j)} < \theta\ .
\end{equation}
\end{itemize}

We consider transmission over the BEC and set the target bit erasure probability $P\stxt{b,max}$ to $10^{-5}$. The number of updates per node $I\stxt{eff}$ (for both check and variable nodes, including the chains and bridges), averaged over the node positions, is considered as a measure of decoding complexity.

\begin{figure}[h]
\setlength{\unitlength}{1mm}
   \begin{picture}(80,105)
   \put(-15,0){\scalebox{1.2}{\includegraphics{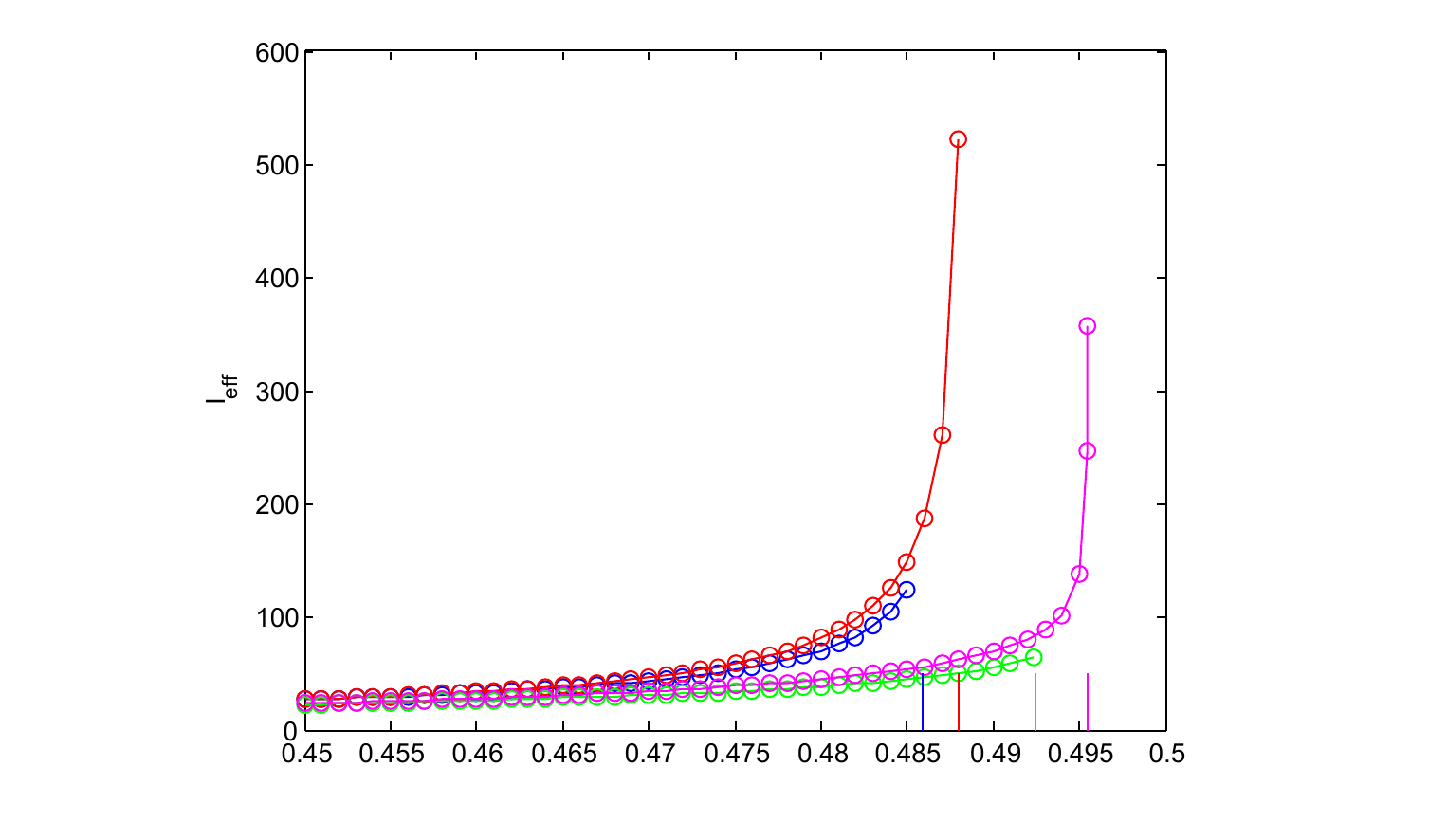}}}
   \put(80,2){$\epsilon$}
\end{picture}
\caption{The average number of updates per node $I\stxt{eff}$ as a function of the BEC parameter $\epsilon$ for the $\mathcal{S}(3,6,24)$ ensemble (green and magenta curves) and the $\mathcal{C}(3,6,18)$ ensemble (blue and red curves). The green and blue curves are computed for the updating schedule with improvement parameter $\theta=10^{-2}$, while the red and magenta curves are for $\theta=0$. The corresponding thresholds are given by vertical lines.}
\label{Fig:Ieff}
\end{figure}

The average number of updates per node $I\stxt{eff}$ is plotted in Fig.~\ref{Fig:Ieff} as a function of the BEC parameter $\epsilon$. The green and magenta curves correspond to the proposed ensemble $\mathcal{S}(3,6,24)$, while the blue and red curves are computed for a single  chain ensemble of length $L=18$, i.e., ensemble $\mathcal{C}(3,6,18)$.\footnote{Ensembles $\mathcal{S}(3,6,24)$ and $\mathcal{C}(3,6,18)$ have been selected for comparison since they both have design rates approximately equal to $0.444$.} The green and blue curves correspond to the updating schedule with the improvement parameter $\theta=10^{-2}$, while the red and magenta curves correspond to $\theta=0$ (in which case, updates are performed regardless of the potential erasure probability improvement).  We  observe a significant complexity improvement provided by the connected chain construction. 
 The vertical straight lines indicate the iterative decoding thresholds calculated for each construction with the corresponding update schedule. Finally, we note that the ensemble $\mathcal{S}(3,6,24)$ also has better thresholds than $\mathcal{C}(3,6,18)$. 

\subsection{Simulation Results}
\label{sec:sims}

In the previous sections we have demonstrated that the connected chain ensembles have superior asymptotic decoding performance when compared to the single chain ensembles. In this section we will show that the connected chain structure also translates into improved decoding performance for finite code lengths. We will examine the  finite length performance of the connected chain ensembles used for transmission over the AWGN channel. 

\begin{figure}[h]
\setlength{\unitlength}{1mm}
   \begin{picture}(155,100)
   \put(0,0){\includegraphics{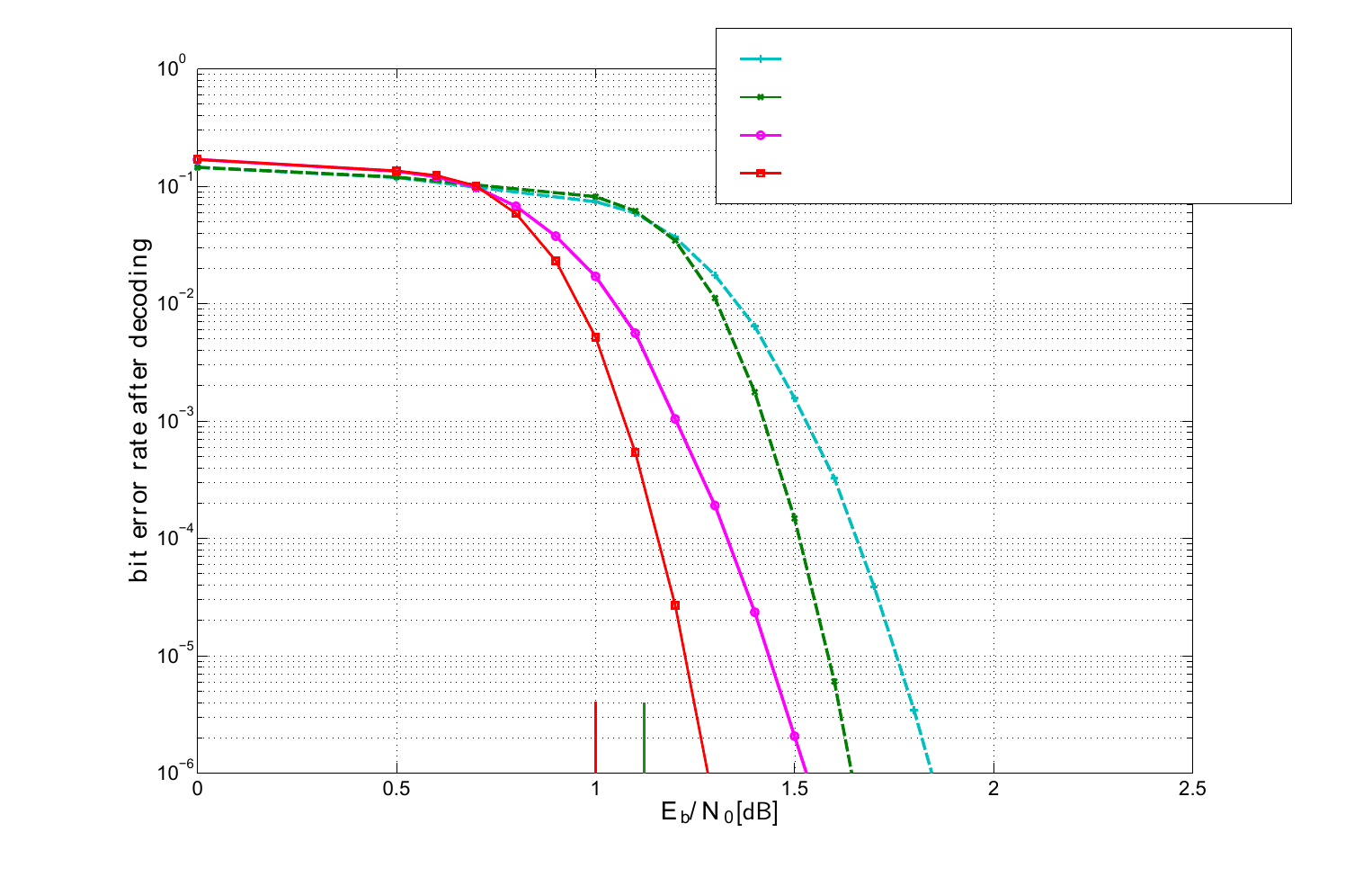}}
   \put(90,92.5){\small $\mathcal{C}(3,6,8),\ n=8192,\ R=0.3752$\normalsize} 
   \put(90,88){\small $\mathcal{C}(3,6,8),\ n=16384,\ R=0.3751$\normalsize} 
   \put(90,83.5){\small $\mathcal{L}(3,6,8),\ n=8192,\ R=0.3755$\normalsize} 
   \put(90,79){\small $\mathcal{L}(3,6,8),\ n=16384,\ R=0.3752$\normalsize} 
\end{picture}
\caption{Bit error rates on the AWGN channel for codes chosen from the $\mathcal{L}(3,6,8)$ ensembles with $M=256$ (purple curve) and $M=512$ (red curve) as well as for codes chosen  from the $\mathcal{C}(3,6,8)$ ensembles with $M=512$ (dashed cyan curve) and $M=1024$ (dashed green curve).}
\label{Fig:loop8_sim_AWGN}
\end{figure}
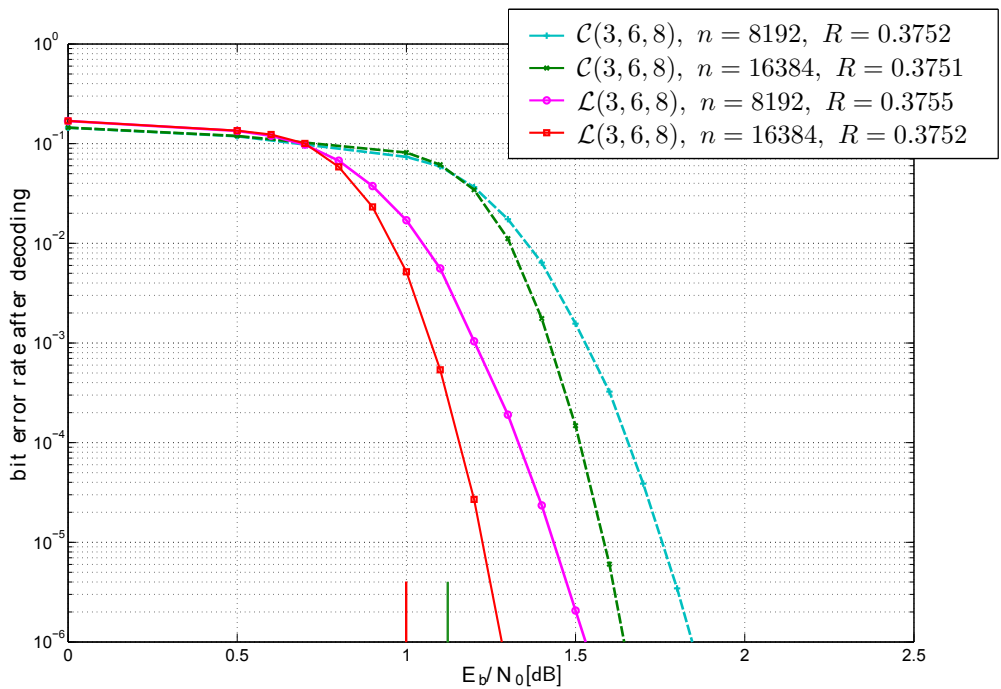

We consider two codes, one of length $n=8192$ and the other of length $n=16384$, randomly selected from the single chain ensemble $\mathcal{C}(3,6,8)$ with lifting factors $M=512$ and $M=1024$, respectively. In addition we randomly pick two codes from the loop $\mathcal{L}(3,6,8)$ ensemble, one with lifting factor $M=256$ and the other with $M=512$. The corresponding code lengths are also $n=8192$ and $n=16384$, respectively. The only condition imposed on the Tanner graphs of the selected codes  was  the absence of cycles of length four. The rate of all codes approximately equals $R=0.375$. The BERs  for transmission over the AWGN channel  are plotted in Fig.~\ref{Fig:loop8_sim_AWGN} as functions of the channel signal-to-noise ratio (SNR) $E\stxt{b}/N_0$. The red and dashed green curves correspond to codes of length $16384$ from the loop and the single chain ensembles respectively. The BERs for the codes of length $8192$ are given by the purple (loop ensemble) and the dashed cyan (single chain ensemble) curves. The asymptotic iterative decoding thresholds for $\mathcal{C}(3,6,8)$ and $\mathcal{L}(3,6,8)$ ensembles are shown by the green and red bars respectively. 

We note that the loop ensemble codes show better decoding performance and deliver a gain of approximately 0.4 dB with respect to the single chain ensemble codes. This happens despite the fact that the loop codes have smaller lifting factors $M$. Similar behavior has been demonstrated  in~\cite{omtc13}, where the loop and  single chain ensemble codes were compared for transmission over the BEC, and the bit erasure rate curves were derived using an analytical approximation.

\subsection{Minimum Distance Analysis}
\label{sec:distance}

In \cite{div06}, Divsalar presented a technique to calculate the average weight enumerator for protograph-based block code ensembles. This weight enumerator can be used to test if an ensemble is \emph{asymptotically good}, i.e., if the minimum distance typical of most members of the ensemble is at least as large as $\delta\stxt{min}n$, where the constant $\delta\stxt{min}>0$ is the \emph{minimum distance growth rate} of the ensemble and $n$ is the block length. In \cite{lmfc10}, it was shown that ensembles of $(J,K)$-regular single chain ensembles are asymptotically good. In this section we present the results of  a similar protograph-based analysis for connected chain ensembles and demonstrate that they share the good distance properties of the individual chains.

Minimum distance growth rates for several square ensembles $\mathcal{S}(3,6,L)$ are shown in Table~\ref{tab:dist}. We observe that, like the individual chain  ensembles, the $\mathcal{S}(3,6,L)$ ensembles are asymptotically good. As the length of the vertical and horizontal chains increases, the rate of the ensemble increases, the iterative decoding thresholds approach the optimum maximum a posteriori probability (MAP) decoding thresholds, and the minimum distance growth rate decreases. This is analogous to the effect of increasing the length $L$ of the single chain ensemble $\mathcal{C}(3,6,L)$ \cite{lmfc10}. 




\begin{table}[h]
\begin{center}
\scalebox{0.95}{%
\begin{tabular}{|c|c|c||c|c|c|}\hline
$L$ & Rate & $\delta\stxt{min}$ & $L$ & Rate & $\delta\stxt{min}$\\\hline
$8$ & $1/3$ &  $0.0136$ & $16$ & $5/12$ & $0.0054$ \\
$10$ & $11/30 $ &$0.0095$ & $18$ & $23/54$ & $0.0048$ \\
$12$ & $7/18$ & $0.0075$ & $20$ & $13/30$ & $0.0043$ \\
$14$ & $17/42$ & $0.0062$ & $24$ & $4/9$ & $0.0036$ \\
\hline
\end{tabular}}
\end{center}
\caption{Minimum distance growth rates for several square ensembles
$\mathcal{S}(3,6,L)$.}\label{tab:dist}\vspace{-5mm}
\end{table}

The asymptotic minimum distance growth rates computed for the ensembles $\mathcal{L}(3,6,L)$ are given in Table~\ref{tab:dist2}. We again observe that the loop ensembles are asymptotically good and note that the growth rates also decrease as the loop lengths, and correspondingly the ensemble rates, increase. 

\begin{table}[h]
\begin{center}
\scalebox{0.9}{%
\begin{tabular}{|c|c|c|}\hline
$L$ & Rate & $\delta\stxt{min}$ \\\hline
$12$ & $0.4167$ &  $0.0109$  \\
$15$ & $0.4333$ &  $0.0085$  \\
$18$ & $0.4444$ & $0.0071$  \\
\hline
\end{tabular}}
\end{center}
\caption{Minimum distance growth rates for several loop ensembles
$\mathcal{L}(3,6,L)$ ensembles.}\label{tab:dist2}\vspace{-5mm}
\end{table}

\section{Degrees of Freedom in Construction}
\label{sec:dof}

Code construction based on the  connection of coupled chains has numerous degrees of freedom. Arbitrary graphs can be created  from single chains of different lengths connected to each other in various ways at different connection points. In this section we will demonstrate how improved connected chain ensembles can be constructed by optimizing the connection point positions and edge placement, by mixing chains of different graph densities, thresholds, and rates,  and by connecting coupled codes based on more general types of LDPC codes.

\subsection{Connection Point Positions}
\label{sec:point}

The first parameter we would like to focus our attention on is the positions at which the chains are connected to each other. The length  proportions of the resulting geometry play an important role in the iterative decoding convergence behavior and threshold of the code ensemble. 

To illustrate the influence of the connection positions, we consider a set of symmetric $(3,6)$-regular loop ensembles parametrized by $h$, the position at which the beginning of the second chain connects to an inner segment of the first chain. For example, the protograph of  the $\mathcal{L}(3,6,15)$ ensemble depicted in Fig.~\ref{Fig:loop} has $h=5$. Now consider fixing $L=15$,  constructing a set of loops for $h=2,3,\cdots,9$, and  computing their thresholds on the BEC. The results are given in Table~\ref{tab:loopsprop}. Note that the maximum threshold value is achieved for $h=5$ 
and the threshold values depend strongly on the connection point positions.

\begin{table}[h]
\begin{center}
 \begin{tabular}{|c|c|c|c|c|c|c|c|c|}
  \hline $h$ & 2 & 3 & 4 & 5 & 6 & 7 & 8 & 9 \\\hline
$\epsilon^*$ & $0.4952$ & $0.4988$& $0.5036$  & $0.5105$ & $0.5039$ & $0.4979$ & $0.4939$ & $0.4912$\\\hline
 \end{tabular}
\end{center}
\caption{BEC thresholds for $(3,6)$-regular loop ensembles in which two individual chains of length $L=15$ are connected $h$ protograph sections from the end of the first chain.}\label{tab:loopsprop}
\end{table}

\begin{figure}[h]
\setlength{\unitlength}{1mm}
   \begin{picture}(155,30)
  \put(0,0){\includegraphics{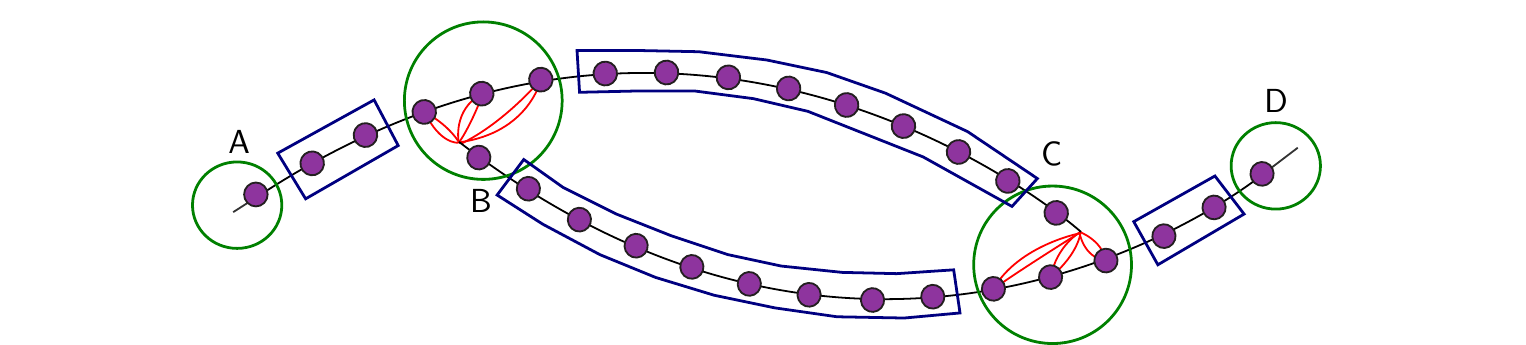}
}
\end{picture}
\caption{Protograph of $\mathcal{L}(3,6,15)$ loop ensemble. The stronger sub-codes A, B, C, and D are denoted by green circles. The intermediate single chain protographs AB, BC, CB, and CD are shown in blue frames.}
\label{fig:looppoints}
\end{figure}

The loop ensemble protograph (see Fig.~\ref{fig:looppoints}) consists of several parts that include two open chain ends  A and D and two connection points B and C, both denoted by green circles, as well as four intermediate single chains AB, BC, CB, and CD, shown by blue frames. The variable nodes in A and D have $3$ outgoing edges each, while the number of outgoing edges per check node is less than $6$. On the other hand, the variable nodes in B and C have more than $3$ outgoing edges per variable node, while each check node has $6$ edges. These eight sections of the protograph, the chain ends, the connection points, and the intermediate chains, which we refer to as sub-graphs, can be regarded as protographs of eight sub-codes which combine to form the code ensemble. At the places where the sub-codes are connected by edges (see Fig.~\ref{fig:looppoints}), messages are exchanged throughout the decoding process. The structure of the sub-codes and their corresponding sub-graphs, as well as their interconnection, determines the decoding behavior of the ensemble.

As noted in Section~\ref{sec:analysis}, the four sections A, B, C, and D form strong sub-codes, which inject reliable information into the  intermediate chains AB, CD, CB, and BC during message passing decoding and initiate decoding convergence across the entire graph. The best choice of connection point locations should result in AB, CD, CB, and BC lengths for which the decoding of these intermediate chains converges simultaneously, so that no chain creates a convergence bottleneck for the entire graph. Therefore, we can conclude that the geometry of a connected chain ensemble should be chosen to ensure simultaneous decoding convergence of the intermediate chains between each pair of strong sub-codes.

\subsection{Forming Connection Points}
\label{sec:connecting}

\begin{figure}[h]
\setlength{\unitlength}{1mm}
   \begin{picture}(155,85)
  \put(0,0){\includegraphics{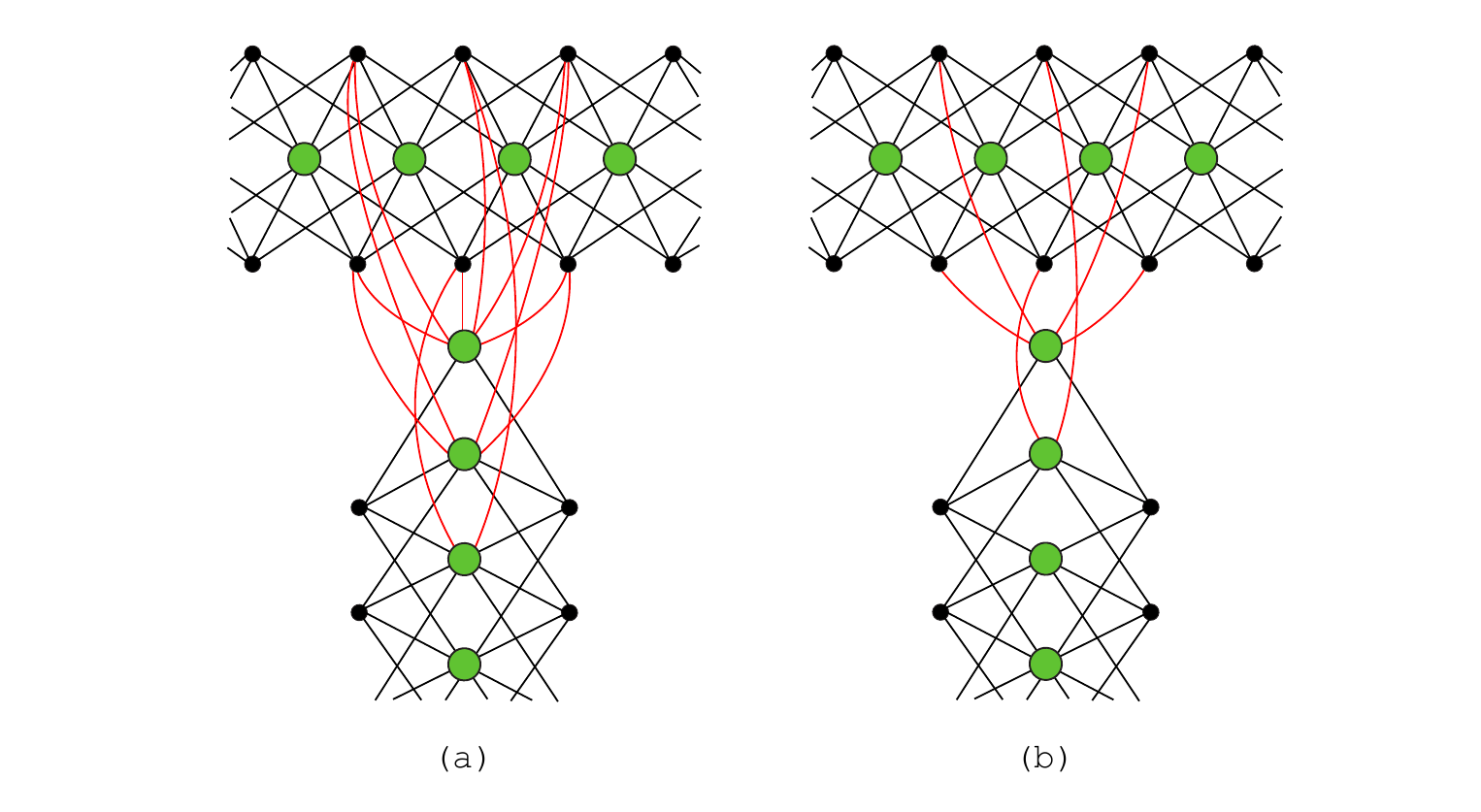}
}
\end{picture}
\caption{Two ways of connecting $(4,8)$-regular protograph chains. The connecting edges are shown in red.}
\label{fig:48conn}
\end{figure}

We now proceed to form a loop consisting of two $(4,8)$-regular chains. 
We consider two different  ways to connect the chains, depicted in Fig.~\ref{fig:48conn}. The first, shown in Fig.~\ref{fig:48conn}(a), has an additional $12$ edges added at the connection point. Here, the check nodes at the connection point each have degree $8$, while the $6$ connecting variable nodes in chain one each have degree $6$. The second connection, depicted in Fig.~\ref{fig:48conn}(b), introduces $6$ additional edges and connects them in a similar fashion to the $(3,6)$-regular chains. As a result, the $3$ check nodes at the connecting end of chain two each have degree $6$, and the $6$ connecting variable nodes of chain one each have degree $5$. We denote the ensembles connected using Fig.~\ref{fig:48conn}(a) by $\mathcal{L}^{A}(4,8,L)$ and the ensembles connected using Fig.~\ref{fig:48conn}(b) by $\mathcal{L}^{B}(4,8,L)$.

BEC thresholds calculated for the ensembles $\mathcal{L}^{A}(4,8,L)$ and $\mathcal{L}^{B}(4,8,L)$ are given in Table~\ref{tab:48thresh}, as well as the thresholds calculated for the single chain ensemble $\chainb$ of the same rate. We see that the thresholds of the ensembles with connection  type B (fewer connecting edges) are always at least as large as the thresholds for connection  type A, with equality occurring for large chain length $L$. We also see the same general behavior as for the $(3,6)$-regular loops: for short $L$ and low rates, the $\mathcal{C}(4,8,L)$ ensemble has the largest threshold; for a large rate region in the middle of the achievable range, the loop ensembles have significantly better thresholds then the single chain ensemble; finally, the improvement observed for the loop ensembles diminishes as $L$ becomes large and the thresholds converge at a value slightly below the MAP threshold of the underlying $(4,8)$-regular LDPC-BC ensemble. The threshold difference between ensembles $\mathcal{L}^{A}(4,8,L)$ and $\mathcal{L}^{B}(4,8,L)$ indicates that the performance is sensitive to the choice of additional edges connecting the chains.

\begin{table}[h]
\begin{center}

\begin{tabular}{|c|c|c|c|c|c|c|c|}\hline
Rate & Ensemble & $\epsilon^*$ & Ensemble & $\epsilon^*$ & Ensemble & $\epsilon^*$ \\\hline
$0.2500$ & $\mathcal{L}^{A}(4,8,6)$ &  $0.5629$ & $\mathcal{L}^{B}(4,8,6)$ & $0.5709$ & $\mathcal{C}(4,8,6)$ & $0.5748$\\
$0.3333$ & $\mathcal{L}^{A}(4,8,9)$ &  $0.5342$ & $\mathcal{L}^{B}(4,8,9)$ & $0.5399$ & $\mathcal{C}(4,8,9)$ & $0.5194$\\
$0.3750$ & $\mathcal{L}^{A}(4,8,12)$ &  $0.5185$ & $\mathcal{L}^{B}(4,8,12)$ & $0.5247$ & $\mathcal{C}(4,8,12)$ & $0.5021$\\
$0.4000$ & $\mathcal{L}^{A}(4,8,15)$ &  $0.5088$ & $\mathcal{L}^{B}(4,8,15)$ & $0.5138$ & $\mathcal{C}(4,8,15)$ & $0.4983$\\
$0.4800$ & $\mathcal{L}^{A}(4,8,75)$ &  $0.4975$ & $\mathcal{L}^{B}(4,8,75)$ & $0.4975$ & $\mathcal{C}(4,8,75)$ & $0.4977$\\
\hline
\end{tabular}
\end{center}
\caption{BEC thresholds for the loop ensembles $\mathcal{L}^{A}(4,8,L)$ and $\mathcal{L}^{B}(4,8,L)$ and the single chain  ensemble $\mathcal{C}(4,8,L)$.}\label{tab:48thresh}
\end{table}

\begin{figure}[h]
\setlength{\unitlength}{1mm}
   \begin{picture}(155,85)
  \put(0,0){\includegraphics{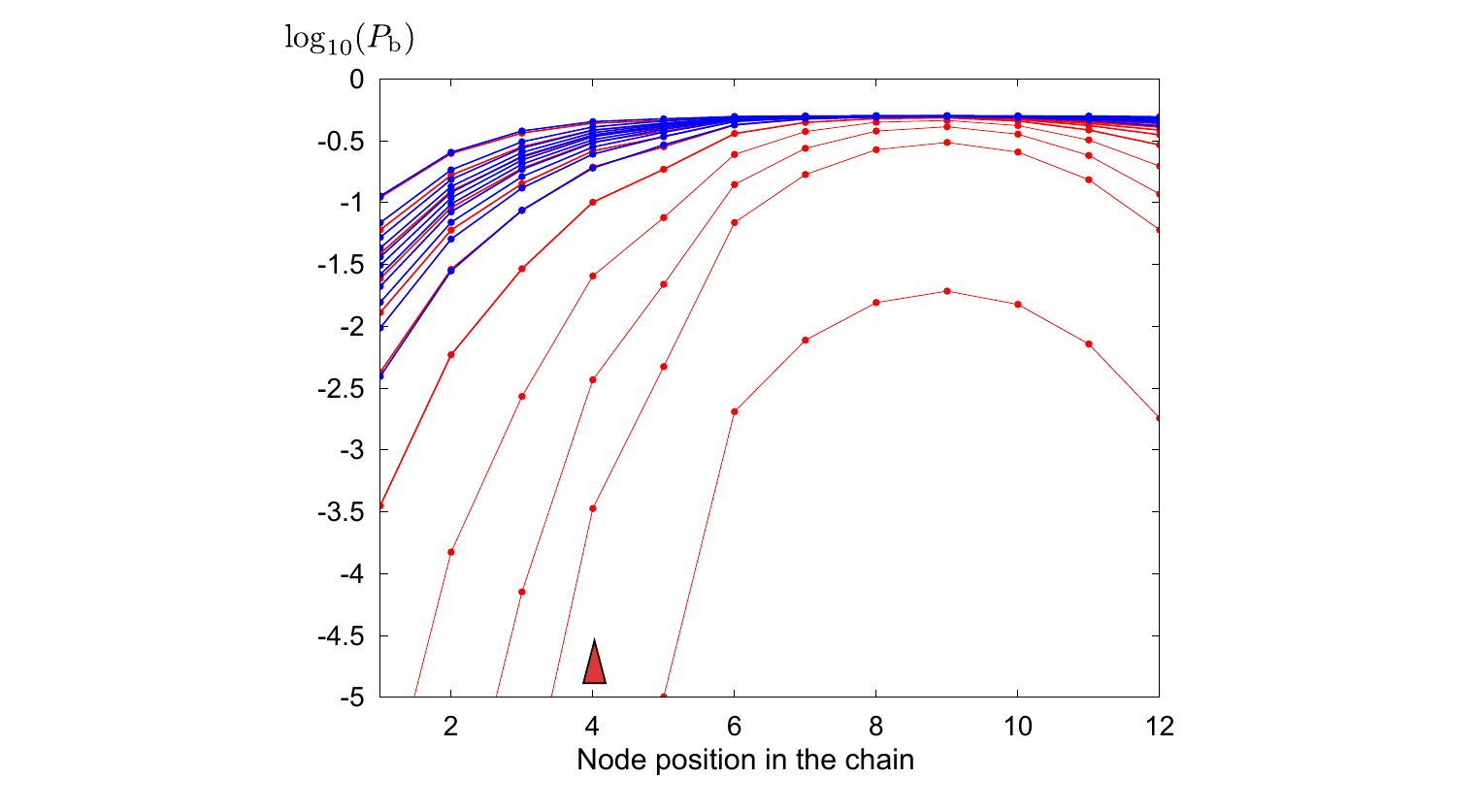}
}
\end{picture}
\caption{Logarithm of the bit erasure probability for the variable nodes of chain one for the ensembles $\mathcal{L}^{A}(4,8,12)$ (blue curves) and $\mathcal{L}^{B}(4,8,12)$  (red curves), as a function of the position of the node in the chain. The curves (either blue or red) are computed for decoding iterations $1,6,11,\ldots,56$ (from top to bottom). The position where chain one is connected to the end of chain two is shown by the red triangle. }
\label{fig:48prob}
\end{figure}

Fig.~\ref{fig:48prob} shows the evolution of the bit erasure probability for the variable nodes of chain one for the  $\mathcal{L}^{A}(4,8,12)$ (blue curves) and $\mathcal{L}^{B}(4,8,12)$ ensembles (red curves). 
The BEC erasure probability is fixed to be $\epsilon=0.514$, and the erasure probabilities are plotted for iteration numbers $1,6,11,\ldots,56$. We observe that the red curves (corresponding to fewer additional edges at the connecting points) achieve lower bit erasure probabilities much faster than the blue curves, i.e., in addition to improved thresholds, connection method B also enables faster convergence to a specific bit erasure probability.

As noted in Section~\ref{sec:connecting}, the convergence of the connected chain ensemble is dictated by the interaction between the stronger sub-codes, formed by the connection points and the chain boundaries, and the weaker sub-codes corresponding to the intermediate chains. Connection types A and B represent two different ways to form the connection point sub-codes. In a type A connection, the variable nodes of the horizontal chain at the connection point are better protected, since they have more outgoing edges than  connection  type B (see Fig.~\ref{fig:48conn}(a)). In a type B connection, the check nodes of the vertical chain at the  connection point are better protected, since they have fewer outgoing edges than connection type A (see Fig.~\ref{fig:48conn}(b)). The results of Table~\ref{tab:48thresh} show that the type B connection is more suitable for the $(4,8)$-regular loop ensemble. We conclude that the connection point structure must be designed to fit the overall convergence schedule of the connected chain ensemble. 

\subsection{Mixing Chains of Different Types}

\begin{figure}[h]
\setlength{\unitlength}{1mm}
   \begin{picture}(155,35)
   \put(0,0){\includegraphics{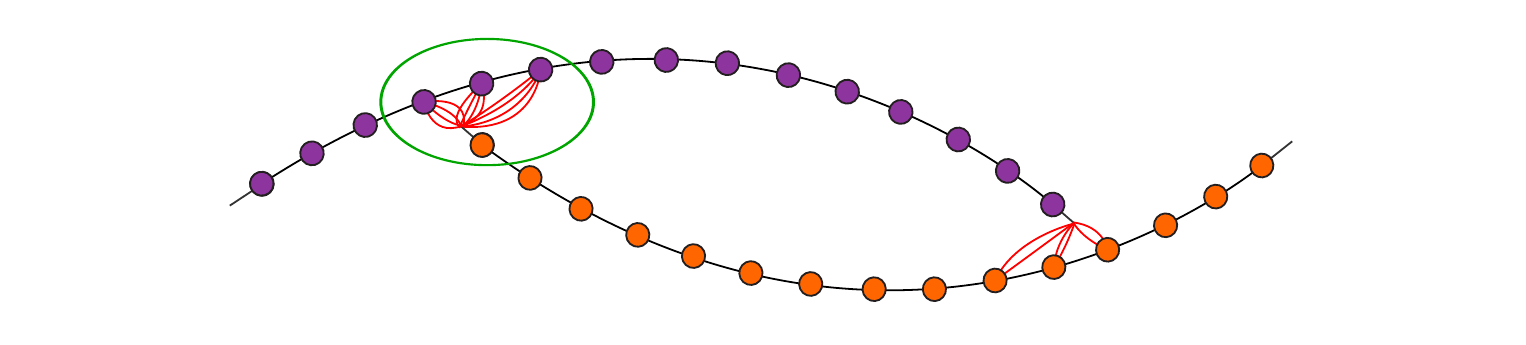}
   }
\end{picture}
\caption{A $(3,6)$-regular protograph chain of length $L=15$  (magenta) is connected to a $(4,8)$-regular protograph chain (orange)  of the same length to form the  $\mathcal{L}_1(3,6,4,8,L)$ ensemble.}
\label{Fig:loop48a36L1}
\end{figure}

\begin{figure}[h]
\setlength{\unitlength}{1mm}
   \begin{picture}(155,80)
   \put(0,0){\includegraphics{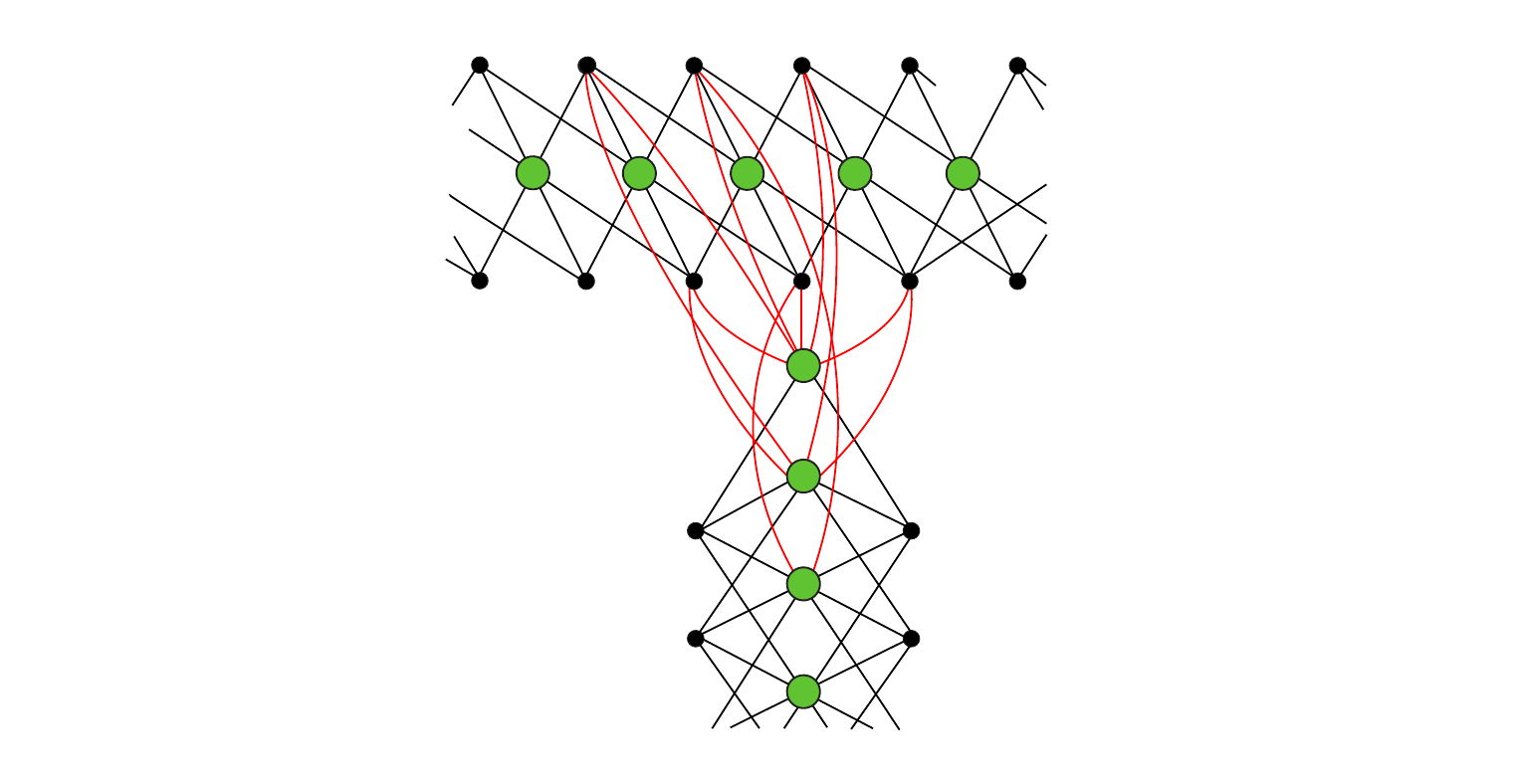}
   }
\end{picture}
\caption{A detailed illustration of the connection between the end of a $(4,8)$-regular chain and the inner segments of the $(3,6)$-regular chain in the $\mathcal{L}_1(3,6,4,8,L)$ ensemble.}
\label{Fig:conn3648}
\end{figure}

Now consider an example of a ``mixed loop'' protograph, obtained by connecting a $(3,6)$-regular chain of length $L$ to a $(4,8)$-regular chain of the same length. We denote this ensemble, illustrated in Fig.~\ref{Fig:loop48a36L1}, by $\mathcal{L}_1(3,6,4,8,L)$. Hereafter, we will refer to the two chains forming the mixed loop as the $(3,6)$ chain and the $(4,8)$ chain. The $(4,8)$ chain segments are shown by orange circles in the figure. The end of each chain is connected to the inner segments of the other chain around the node position $L/3$. The last check node of the $(4,8)$ chain has only 2 outgoing edges, the next to last check node has only 4, and the second to last check node has only 6. Thus there are 12 new edges that can be used to connect the end of the $(4,8)$ chain to the $(3,6)$ chain. The connection pattern, shown inside a green circle in Fig.~\ref{Fig:loop48a36L1}, is detailed in Fig.~\ref{Fig:conn3648}.

\begin{figure}[h]
\setlength{\unitlength}{1mm}
   \begin{picture}(155,35)
   \put(0,0){\includegraphics{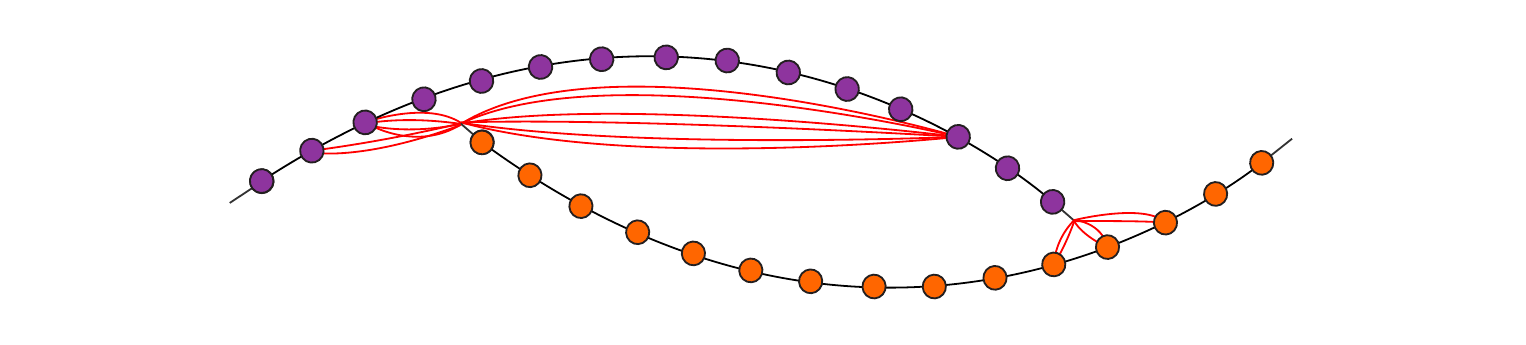}
   }
\end{picture}
\caption{A $(3,6)$ protograph chain of length $L=15$ (magenta) is connected to a $(4,8)$ protograph chain (orange) of the same length to form the $\mathcal{L}_2(3,6,4,8,L)$ ensemble.}
\label{Fig:loop48a36L2}
\end{figure}

We now again demonstrate that the choice of the edge connections is an important design issue.
Consider the alternative mixed loop ensemble $\mathcal{L}_2(3,6,4,8,L)$ presented in Fig.~\ref{Fig:loop48a36L2}. The 12 edges connecting the end of the $(4,8)$ chain to the
$(3,6)$ chain are spread along the $(3,6)$ chain. In particular, the last check node of the $(4,8)$ chain is connected to the nodes at position  13 of the $(3,6)$ chain by 6 edges,
the next to last check node of the $(4,8)$ chain is connected to the nodes at position 3 of the $(3,6)$ chain by 4 edges, and the second to last check node of the $(4,8)$ chain is connected to the nodes at position 2 of the $(3,6)$ chain by 2 edges.

The BEC iterative decoding thresholds of the two mixed loop ensembles are given in Table~\ref{tab:thresmix}. We notice that the thresholds of both the $\mathcal{L}_1(3,6,4,8,15)$
and $\mathcal{L}_2(3,6,4,8,15)$ ensembles are better than the thresholds of single chains of the same rate. On the other hand, the threshold of the ensemble $\mathcal{L}_2(3,6,4,8,15)$
(with optimized connections) is significantly better than the threshold of the ensemble $\mathcal{L}_1(3,6,4,8,15)$, whose construction mimics the $\mathcal{L}(3,6,15)$ loop.
The placement of the connections in $\mathcal{L}_2(3,6,4,8,15)$ takes into account the difference in the behavior of the connected chains. The first 6 connections from the end of the
$(4,8)$ chain connect to one end of the $(3,6)$ chain to help it converge, while the other 6 connections are placed nearly at the other end of the $(3,6)$ chain, where it, in turn, connects to the
$(4,8)$ chain. As a result, the second set of 6 connections helps the convergence of the $(4,8)$ chain. This is important, because the $(4,8)$ chain requires a stronger initial boost to convergence in order to display threshold improvement.


\begin{table}[h]
\begin{center}
\scalebox{0.9}{%
\begin{tabular}{|c|c|}\hline
Ensemble & $\epsilon^*$  \\\hline
$\mathcal{L}_1(3,6,4,8,15)$ &  $0.4997$ \\
$\mathcal{L}_2(3,6,4,8,15)$ &  $0.5105$ \\
$\mathcal{C}(3,6,12)$ & $0.495$\\
$\mathcal{C}(4,8,18)$ & $0.4977$\\
\hline
\end{tabular}}
\end{center}
\caption{BEC thresholds $\epsilon^*$ for mixed loop ensembles and single chain ensembles of  rate $R=0.417$.}\label{tab:thresmix}\vspace{-4mm}
\end{table}

Choosing how to connect chains of different types in general is challenging, since each part of the resulting protograph - a connection point, a boundary point, or an intermediate single chain  -- acts differently and has different convergence properties. The overall design must account for these differences. More specifically, we can distinguish several phases of message passing decoding convergence: the initial phase in which the boundary points and connection points inject reliable information to the graph, the middle phase in which the intermediate chains start to converge, and the final phase in which decoding finishes. When chains of different types are connected, their speed of convergence, their thresholds, and the amount of reliable information required to initiate convergence for each type must be  utilized jointly via the connection geometry so that at each phase iterative decoding convergence spreads  over the entire protograph.

\subsection{Connecting Different Code Ensembles}

In this section we consider connected spatially coupled single chain ensembles built from irregular  LDPC-BC ensembles - the ARJA and AR4JA codes~\cite{ddja09}, which are known to exhibit very good iterative decoding performance. The protographs of these code ensembles are optimized to yield thresholds superior to these of regular LDPC-BC ensembles. The ARJA  and  AR4JA code ensembles also have  linear minimum distance growth, despite the fact that they are irregular.

\begin{figure}[h]
\setlength{\unitlength}{1mm}
   \begin{picture}(155,40)
   \put(0,0){\includegraphics{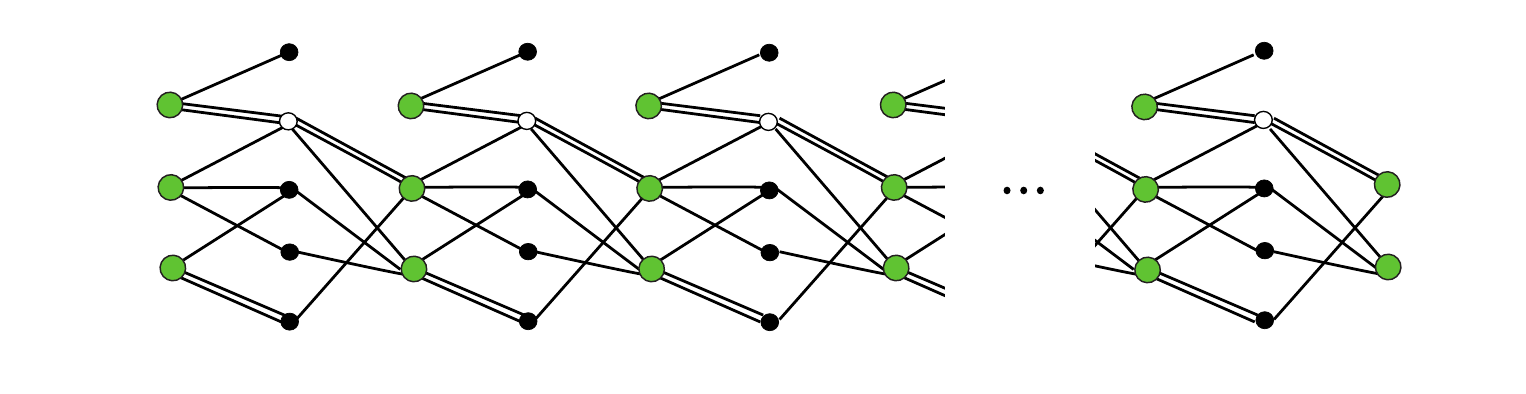}
   }
\end{picture}
\caption{Protograph of a single chain ensemble built from protographs of rate $R=1/2$ ARJA codes. The white circles represent punctured variable nodes.}
\label{Fig:arja_chain}
\end{figure}

Single chain ensembles constructed by connecting ARJA and AJ4JA photographs were considered in~\cite{mlc10}, and it was demonstrated there that spatial coupling improves the BP thresholds of these codes. The protograph of a single chain ensemble built from rate $R=1/2$ ARJA protographs is shown in Fig.~\ref{Fig:arja_chain}. We now show that  connecting two single chain ARJA ensembles can further improve their decoding performance.

\begin{figure}[h]
\setlength{\unitlength}{1mm}
   \begin{picture}(155,130)
   \put(0,0){\includegraphics{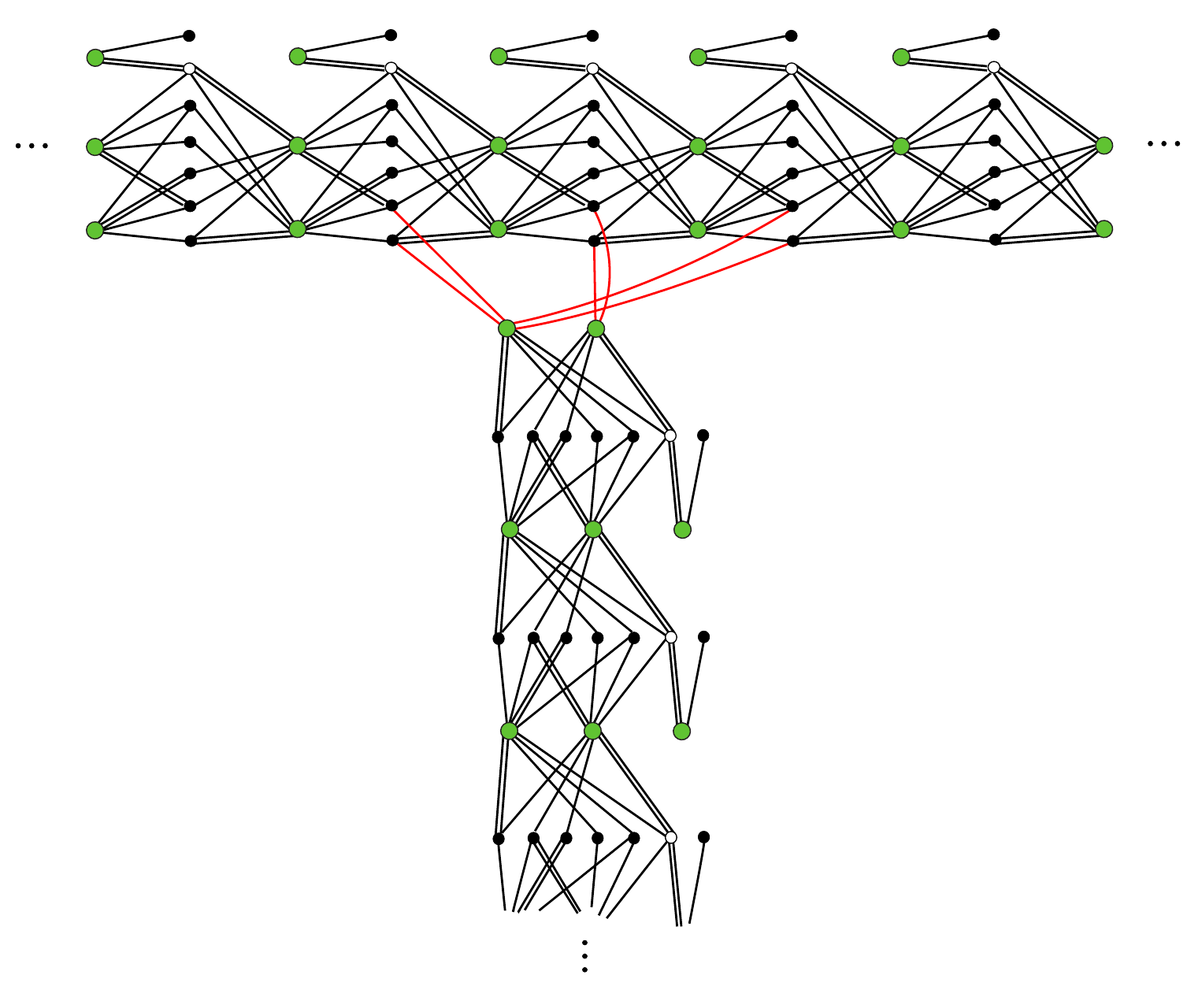}
   }
\end{picture}
\caption{Example connection of two AR4JA chains.}
\label{Fig:ar4ja_con}
\end{figure}

\begin{figure}[h]
\setlength{\unitlength}{1mm}
   \begin{picture}(155,100)
   \put(0,0){\includegraphics{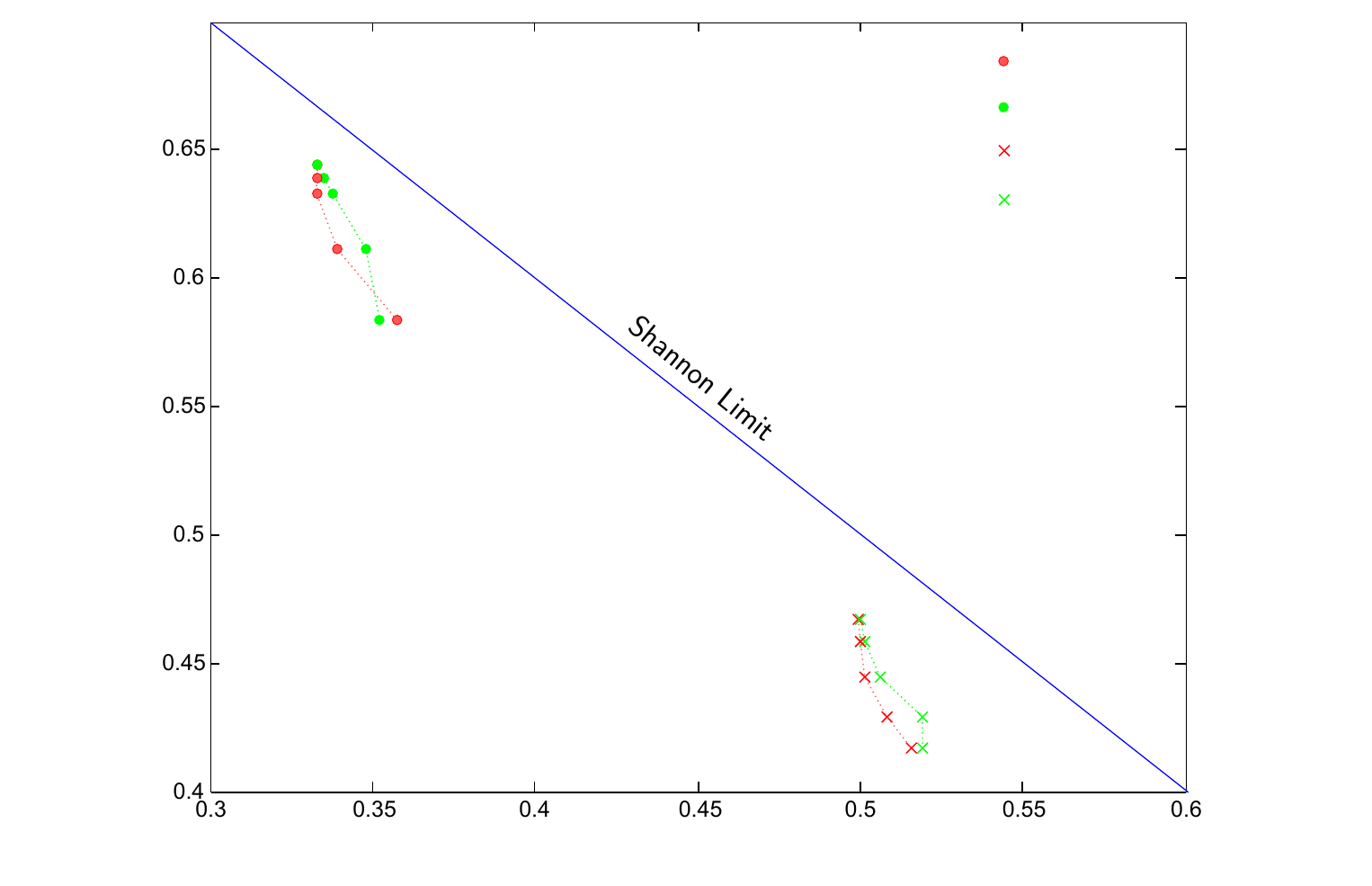}}
   \put(77,2){$\epsilon^*$} 
   \put(12,52){$R$}
   \put(116,92){\small $\mathcal{C}\stxt{AR4JA}(L)$}
   \put(116,87){\small $\mathcal{L}\stxt{AR4JA}(L)$}
   \put(116,82){\small $\mathcal{C}\stxt{ARJA}(L)$}
   \put(116,77){\small $\mathcal{L}\stxt{ARJA}(L)$}
\end{picture}
\caption{BEC thresholds of the single chain ensembles $\mathcal{C}\stxt{ARJA}(L)$ and $\mathcal{C}\stxt{AR4JA}(L)$ (red crosses and dots) compared to the thresholds of the loop ensembles $\mathcal{L}\stxt{ARJA}(L)$ and $\mathcal{L}\stxt{AR4JA}(L)$ (green crosses and dots).}
\label{Fig:arja_thresholds}
\end{figure}

We consider both the single chain ARJA ensembles, denoted by $\mathcal{C}\stxt{ARJA}(L)$, where we connect $L$ ARJA protographs (see Fig.~\ref{Fig:arja_chain}) of rate $R=1/2$, and the single chain AR4JA ensembles,  denoted by $\mathcal{C}\stxt{AR4JA}(L)$, where $L$ protographs of AR4JA ensembles of rate $R=2/3$ are connected. We compare these to the loop ensembles $\mathcal{L}\stxt{ARJA}(L)$ constructed by connecting two $\mathcal{C}\stxt{ARJA}(L)$ chains to make a loop with proportions $L/3$ and $2L/3$, as in Fig.~\ref{Fig:loop}, 
and the loop ensembles $\mathcal{L}\stxt{AR4JA}(L)$, constructed from two AR4JA chains using the same technique. Fig.~\ref{Fig:ar4ja_con} shows an example of connecting two AR4JA chains.

Fig.~\ref{Fig:arja_thresholds} depicts the BP thresholds of the loop and single chain ARJA and AR4JA ensembles on the BEC as functions of the ensemble rate $R$. The red crosses correspond to the single chain ensembles $\mathcal{C}\stxt{ARJA}(L)$ with $L=6,7,9,12$, and $15$ (from bottom to top) and the green crosses correspond to the loop ensembles $\mathcal{L}\stxt{ARJA}(L)$ with the same $L$. The red dots correspond to the single chain ensembles $\mathcal{C}\stxt{AR4JA}(L)$ with $L=4,6,10,12$, and $15$ (from bottom to top) and the green dots correspond to the loop ensembles $\mathcal{L}\stxt{AR4JA}(L)$ with the same $L$. Note that, for both cases, the loop ensembles have better thresholds for an intermediate range of values of $L$. 

We observe that the chain connection principle successfully applies to irregular  code chains and that  performance improvement is obtained. On the other hand, we notice that, in general, connecting irregular protograph chains requires a greater optimization effort to yield the best result, since irregular protograph chains have more degrees of freedom. Analysis techniques such as density evolution or bit erasure  probability evolution charts (see Figs.~\ref{Fig:Pb_chain}~and~\ref{fig:48prob}), which relate the protograph structure to the decoding convergence behavior,  can be utilized as design guidelines.

\section{Conclusions}
\label{sec:conc}

The connection of spatially coupled protograph chains provides an approach to extend the spatial
graph coupling phenomenon from simple (single chain) graph coupling to more general coupled structures. We have shown that connecting spatially coupled chains results in protograph-based LDPC code ensembles with improved thresholds and reduced iterative decoding complexity. Simulation results demonstrate that the asymptotic threshold and complexity improvements translate into improved finite-length performance. Moreover, connected chain ensembles are asymptotically good in terms of minimum distance. 

There are many possible variations on the basic construction method. Our results indicate that performance is sensitive to various parameters, such as the distance between connection points, the placement of connecting edges, and the individual characteristics of the component chains, and we provided insight into the relation between these parameters and the resulting decoding performance. 
Finally, we note that the principle of coupled chain connection is very general and may give rise to many novel LDPC code ensemble constructions.



\bibliographystyle{IEEEtran}


\begin{thebibliography}{10}
\providecommand{\url}[1]{#1}
\csname url@rmstyle\endcsname
\providecommand{\newblock}{\relax}
\providecommand{\bibinfo}[2]{#2}
\providecommand\BIBentrySTDinterwordspacing{\spaceskip=0pt\relax}
\providecommand\BIBentryALTinterwordstretchfactor{4}
\providecommand\BIBentryALTinterwordspacing{\spaceskip=\fontdimen2\font plus
\BIBentryALTinterwordstretchfactor\fontdimen3\font minus
  \fontdimen4\font\relax}
\providecommand\BIBforeignlanguage[2]{{%
\expandafter\ifx\csname l@#1\endcsname\relax
\typeout{** WARNING: IEEEtran.bst: No hyphenation pattern has been}%
\typeout{** loaded for the language `#1'. Using the pattern for}%
\typeout{** the default language instead.}%
\else
\language=\csname l@#1\endcsname
\fi
#2}}

\bibitem{Gal63}
R.~G. Gallager, ``Low-density parity-check codes,'' Ph.D. dissertation,
  Massachusetts Institute of Technology, Cambridge, MA, 1963.

\bibitem{mac98}
D.~J.~C. MacKay, ``Gallager codes that are better than turbo codes,'' in
  \emph{Proc. Thirty-sixth Allerton Conference on Communication, Control, and
  Computing}, Monticello, IL, Sept. 1998.

\bibitem{mac99}
------, ``Good error-correcting codes based on very sparse matrices,''
  \emph{IEEE Transactions on Information Theory}, vol.~45, no.~2, pp. 399--431,
  Mar. 1999.

\bibitem{ru01b}
T.~J. Richardson and R.~L. Urbanke, ``The capacity of low-density parity-check
  codes under message-passing decoding,'' \emph{IEEE Transactions on
  Information Theory}, vol.~47, no.~2, pp. 599--618, Feb. 2001.

\bibitem{fz99}
A.~{Jim\'{e}nez Felstr\"{o}m} and {K. Sh. Zigangirov}, ``Time-varying periodic
  convolutional codes with low-density parity-check matrices,'' \emph{IEEE
  Transactions on Information Theory}, vol.~45, no.~6, pp. 2181--2191, Sept.
  1999.

\bibitem{lscz10}
M.~Lentmaier, A.~Sridharan, D.~J. {Costello, Jr.}, and {K. Sh. Zigangirov},
  ``Iterative decoding threshold analysis for {LDPC} convolutional codes,''
  \emph{IEEE Transactions on Information Theory}, vol.~56, no.~10, pp.
  5274--5289, Oct. 2010.

\bibitem{kru11}
S.~Kudekar, T.~J. Richardson, and R.~L. Urbanke, ``Threshold saturation via
  spatial coupling: why convolutional {LDPC} ensembles perform so well over the
  {BEC},'' \emph{IEEE Transactions on Information Theory}, vol.~57, no.~2, pp.
  803--834, Feb. 2011.

\bibitem{kru12}
\BIBentryALTinterwordspacing
S.~Kudekar, T.~Richardson, and R.~Urbanke, ``Spatially coupled ensembles
  universally achieve capacity under belief propagation,'' 2012. [Online].
  Available: \url{http://arxiv.org/abs/1201.2999}
\BIBentrySTDinterwordspacing

\bibitem{kymp12}
S.~Kumar, A.~J. Young, N.~Macris, and H.~D. Pfister, ``A proof of threshold
  saturation for spatially-coupled {LDPC} codes on {BMS} channels,'' in
  \emph{Proc. Fiftieth Annual Allerton Conference}, Monticello, IL, Oct. 2012.

\bibitem{kp10}
S.~Kudekar and H.~D. Pfister, ``The effect of spatial coupling on compressive
  sensing,'' in \emph{Proc. Allerton Conference on Communications, Control, and
  Computing}, Monticello, IL, Sept. 2010.

\bibitem{tru12comlet}
D.~Truhachev, ``Achieving {AWGN} multiple access channel capacity with spatial
  graph coupling,'' \emph{IEEE Communications Letters}, vol.~16, no.~5, pp.
  585--588, May 2012.

\bibitem{tru13isit}
------, ``Universal multiple access via spatially coupling data transmission,''
  in \emph{Proc. IEEE International Symposium on Information Theory}, Istanbul,
  Turkey, July 2013.

\bibitem{ttk11}
K.~Takeuchi, T.~Tanaka, and T.~Kawabata, ``Improvement of {BP}-based {CDMA}
  multiuser detection by spatial coupling,'' in \emph{Proc. IEEE International
  Symposium on Information Theory}, St. Petersburg, Russia, Aug. 2011.

\bibitem{hkis11}
M.~Hagiwara, K.~Kasai, H.~Imai, and K.~Sakaniwa, ``Spatially coupled
  quasi-cyclic quantum {LDPC} codes,'' in \emph{Proc. IEEE International
  Symposium on Information Theory}, St. Petersburg, Russia, Aug. 2011.

\bibitem{sts11}
Z.~Si, R.~Thobaben, and M.~Skoglund, ``Bilayer {LDPC} convolutional codes for
  half-duplex relay channels,'' in \emph{Proc. IEEE International Symposium on
  Information Theory}, St. Petersburg, Russia, Aug. 2011.

\bibitem{ruas11}
V.~Rathi, R.~Urbanke, M.~Andersson, and M.~Skoglund, ``Rate-equivocation
  optimally spatially coupled {LDPC} codes for the {BEC} wiretap channel,'' in
  \emph{Proc. IEEE International Symposium on Information Theory}, St.
  Petersburg, Russia, Aug. 2011.

\bibitem{ypn11}
A.~Yedla, H.~Pfister, and K.~Narayanan, ``Universality for the noisy
  {Slepian-Wolf} problem via spatial coupling,'' in \emph{Proc. IEEE
  International Symposium on Information Theory}, St. Petersburg, Russia, Aug.
  2011.

\bibitem{au11}
V.~Aref and R.~Urbanke, ``Universal rateless codes from coupled {LT} codes,''
  in \emph{Proc. IEEE Information Theory Workshop}, Paraty, Brazil, Oct. 2011.

\bibitem{ddja09}
D.~Divsalar, S.~Dolinar, C.~Jones, and K.~Andrews, ``Capacity-approaching
  protograph codes,'' \emph{IEEE Journal on Selected Areas in Communications},
  vol.~27, no.~6, pp. 876--888, Aug. 2009.

\bibitem{tho03}
J.~Thorpe, ``Low-density parity-check ({LDPC}) codes constructed from
  protographs,'' Jet Propulsion Laboratory, Pasadena, CA, INP Progress Report
  42-154, Aug. 2003.

\bibitem{lpf11}
M.~Lentmaier, M.~M. Prenda, and G.~Fettweis, ``Efficient message passing
  scheduling for terminated {LDPC} convolutional codes,'' in \emph{Proc. IEEE
  International Symposium on Information Theory}, St. Petersburg, Russia, Aug.
  2011.

\bibitem{omtc13}
P.~M. Olmos, D.~G.~M. Mitchell, D.~Truhachev, and {D. J. Costello, Jr.}, ``A
  finite length performance analysis of {LDPC} codes constructed by connecting
  spatially coupled chains,'' in \emph{Proc. IEEE Information Theory Workshop},
  Seville, Spain, Sept. 2013.

\bibitem{div06}
D.~Divsalar, ``Ensemble weight enumerators for protograph {LDPC} codes,'' in
  \emph{Proc. IEEE International Symposium on Information Theory}, Seattle, WA,
  July 2006.

\bibitem{lmfc10}
M.~Lentmaier, D.~G.~M. Mitchell, G.~P. Fettweis, and D.~J. {Costello, Jr.},
  ``Asymptotically regular {LDPC} codes with linear distance growth and
  thresholds close to capacity,'' in \emph{Proc. Information Theory and
  Applications Workshop}, San Diego, CA, Feb. 2010.

\bibitem{mlc10}
D.~G.~M. Mitchell, M.~Lentmaier, and D.~J. {Costello, Jr.}, ``New families of
  {LDPC} block codes formed by terminating irregular protograph-based {LDPC}
  convolutional codes,'' in \emph{Proc. IEEE International Symposium on
  Information Theory}, Austin, TX, June 2010.

\end{thebibliography}

\end{document}